\documentclass[aps,prl,twocolumn,showpacs,superscriptaddress,nobibnotes,nofootinbib,showkeys,10pt]{revtex4-2} 
\usepackage[dvips]{graphicx}
\usepackage{epsfig}
\usepackage{bm}   
\usepackage{dcolumn}
\usepackage{rotate}
\usepackage[table]{xcolor}
\usepackage{footmisc,booktabs,amssymb,longtable}
\usepackage{amsmath,amsfonts}

\usepackage{multirow}
\definecolor{Gray}{gray}{0.85}
\definecolor{LightCyan}{rgb}{0.88,1,1}

\usepackage[scaled=0.92]{helvet}
\usepackage[T1]{fontenc}
\usepackage{textcomp}
\usepackage{braket}
\usepackage{textgreek}
\usepackage{rotate}
\usepackage{tabularx}
\usepackage[referable]{threeparttablex}

\definecolor{capri}{rgb}{0.0, 0.75, 1.0}
\definecolor{cornflowerblue}{rgb}{0.39, 0.58, 0.93}
\definecolor{spirodiscoball}{rgb}{0.06, 0.75, 0.99}
\definecolor{pear}{rgb}{0.82, 0.89, 0.19}

\begin{document}

\title[Electric dipole polarizability of low-lying excited states in atomic nuclei]{Electric dipole polarizability of low-lying excited states in atomic nuclei}

\author{Jos\'e Nicol\'as Orce}
\thanks{Email: coulex@gmail.com}   \email{https://github.com/UWCNuclear/} \homepage{http://nuclear.uwc.ac.za}
\affiliation{Department of Physics \& Astronomy, University of the Western Cape, P/B X17, Bellville 7535, South Africa}
\affiliation{National Institute for Theoretical and Computational Sciences (NITheCS), South Africa}

\author{Cebo Ngwetsheni}
\affiliation{Department of Physics \& Astronomy, University of the Western Cape, P/B X17, Bellville 7535, South Africa}

\date{\today}

\begin{abstract}

Novel equations for the electric dipole  polarizability $\alpha_{_{E1}}$ of
low-lying excited states in atomic nuclei --- and the
related $(-2)$ moment of the total photo-absorption cross section, $\sigma_{_{-2}}$ ---  are inferred in terms of electric dipole and quadrupole matrix elements.
These equations are valid for arbitrary angular momenta of the initial/ground and final/excited states and have been exploited in
fully converged 1$\hbar\omega$ shell-model calculations of selected {\it p-} and {\it sd-}shell nuclei that consider configuration mixing;
advancing previous knowledge from $^{17}$O to $^{36}$Ar, where thousands of electric dipole matrix elements are computed from isovector excitations which include  the giant dipole resonance region.
Our results are in reasonable agreement with previous shell-model calculations
and follow --- except for $^{6,7}$Li  and $^{17,18}$O --- Migdal's global trend provided by the combination of the hydrodynamic model and second-order non-degenerate
perturbation theory. Discrepancies in $^{6,7}$Li  and $^{17}$O arise as a result of  the presence of $\alpha$-cluster configurations in odd-mass nuclei,
whereas the disagreement in $^{18}$O comes from the mixing of intruder states, which is lacking in the shell-model interactions.
More advanced \emph{ab initio} calculations of the dipole polarizability for low-lying excited states covering all the isovector states within the giant dipole resonance region are missing and could be very valuable to benchmark the results presented here and shed further light on how atomic nuclei polarize away from the ground state


\end{abstract}



\maketitle

\section{Motivation}

The bulk of knowledge on the nuclear electric dipole ({\sc E1}) polarizability,  $\alpha_{_{E1}}$, concerns the ground states of nuclei and arises from photo-absorption cross-section data~\cite{dietrich1988atlas,plujko2018giant,kawano2020iaea}, where most of the absorption (and emission) of  photons is provided by the giant dipole resonance ({\sc GDR})~\cite{ishkhanov2021giant}. The latter is understood macroscopically as the collective motion of inter-penetrating proton and neutron fluids out of phase~\cite{migdal1945quadrupole,goldhaber1948nuclear,steinwedel1950nuclear}, whereas  is described microscopically by the shell-model ({\sc SM}) interpretation of a system of independent nucleons or particle-hole excitations plus configuration mixing~\cite{levinger1954independent,balashov1962relation,danos1965photonuclear}.
Data predominantly involve photo-neutron cross sections, although photo-proton contributions are relevant for light and $N=Z$ self-conjugate nuclei~\cite{orce2022competition}.
To a much lesser extend,  $\alpha_{_{E1}}$ has  been determined from several experiments using radioactive ion beams~\cite{rossi2013measurement}, inelastic proton scattering~\cite{tamii2011complete,roca2015neutron,hashimoto2015dipole,roca2018nuclear,bassauer2020evolution} and virtual photons~\cite{orce2020polarizability}.
The latter are also responsible for the polarization of atoms and molecules~\cite{mott1948}.

The understanding of how $\alpha_{_{E1}}$ evolves as a function of excitation energy is relevant for nuclear structure physics~\cite{orce2020polarizability} and nuclear astrophysics~\cite{orce2021enhanced}.
Average properties can be extracted from {\sc GDR}s built on excited states
by fitting the {\sc GDR} energy and width parameters to data~\cite{orce2021enhanced}, using the second-sound hydrodynamic model~\cite{danos1958long} and assuming the validity of the  Brink-Axel hypothesis~\cite{brink_1955,axel1962electric}.
The latter seems validated below critical temperatures of $T\lessapprox T_c = 0.7 +37.5/A$ MeV and angular momenta
$J \lessapprox J_c=0.6 A^{5/6}$~\cite{kusnezov1998scaling}, where excited {\sc GDR}s present similar parameters
to their ground-state counterparts~\cite{snover1986giant,gaardhoje1992nuclear}.
There is bountiful information  for $T\gg$1 MeV from heavy-ion fusion-evaporation reactions~\cite{snover1986giant,gaardhoje1992nuclear,schiller2007compilation,bracco1998status}, some in the range $0.7\lessapprox T\lessapprox1$ MeV~\cite{mondal2018study,kicinska1987statistical,dey2014probing,heckman2003low,pandit2012critical,gossett1985deformation, pandit2021puzzle}, and hardly anything for $0 \lessapprox T < 0.7$ MeV.
At $T\approx0$ MeV, $\alpha_{_{E1}}$ has been determined  from Coulomb-excitation reactions for only a couple of favorable cases with excited states $J=1/2$  ---
$^7$Li~\cite{hausser1973e1,vermeer1984coulomb,vermeer1984experimental} and $^{17}$O~\cite{kuehner1982measurement} --- where the spectroscopic or static quadrupole moment is zero, $Q_{_S}(J=1/2)=0$\footnote{From the vanishing $3j$ Wigner symbol $\propto J(2J-1)$~\cite{alder1975north,deshalit1974theoretical,suhonen2007nucleons}.}.
Inasmuch as $\alpha_{_{E1}}$, $Q_{_S}(J)$ is a second-order effect in Coulomb-excitation perturbation theory~\cite{alder1975north,eichler1964second,deBoer1968,hausser1974} that provides a measure of
the extent to which the nuclear charge distribution in the laboratory frame acquires an ellipsoidal deformation.
The empirical disentanglement of $\alpha_{_{E1}}$ and $Q_{_S}$ values for excited states with  $J\neq1/2$ requires increasing experimental accuracy and has never been done.

For ground states, $\alpha_{_{E1}}$ can be deduced using non-degenerate perturbation theory by means of the energy-shift of nuclear levels arising from the
quadratic Stark effect~\cite{flambaum2021nuclear}, and has been investigated with {\sc SM} calculations~\cite{barger1982gdr,barker1982gdr17O,barker1982gdr18O,barke1989investigation,raju2018reorientation,orce2023global} using,
\begin{eqnarray}
\alpha_{_{E1}}
&=& \frac{2e^2}{2J_i+1} \sum_n \frac{\left|\langle i\parallel\hat{E1}\parallel
n\rangle\right|^2}{E_n - E_i} = \frac{9\hbar c}{8\pi^3}\sigma_{_{-2}},
\label{eq:polar2}
\end{eqnarray}
where the sum extend over $\lvert n\rangle$ intermediate states
connecting  the initial/ground state $\lvert i\rangle$ with isovector $E1$ transitions~\cite{warburton1969role},
$2J_i+1$ is the normalization constant arising from the Wigner-Eckart theorem~\cite{rose1957elementary,messiah1961quantum} ---
validating Eq.~\ref{eq:polar2} for arbitrary $J_i$ ground states ---  and $\sigma_{_{-2}}$ the $(-2)$ moment of the total
photo-absorption cross section, $\sigma_{_{total}}(E_{\gamma})$,  defined by~\cite{levinger1960,migdal1965theory},
\begin{eqnarray}
\sigma_{_{-2}}=\int_{S_n}^{E_{\gamma}^{max}}
\frac{\sigma_{_{total}}(E_{{\gamma}})}{E_{{\gamma}}^{^2}}dE_{{\gamma}},
\label{eq:sigma-2int2}
\end{eqnarray}
which is generally integrated between neutron threshold $S_n$ and the experimentally available upper limit for monochromatic photons,
$E_{\gamma}^{max}\approx20-50$ MeV~\cite{dietrich1988atlas}.
An upper
limit of $E_{\gamma}^{max}\approx50$ MeV approximates the $\sigma_{_{-2}}$ asymptotic value for light and medium-mass
nuclei~\cite{ahrens1976experimental}. For heavy nuclei with atomic mass number $A=N+Z\gtrapprox50$, $\sigma_{_{-2}}$ values generally follow the empirical power-law formula~\cite{orce2015new,orce2016reply},
\begin{eqnarray}
\sigma_{_{-2}}(A) = 2.38\kappa ~ A^{5/3}\,\mbox{$\mu$b/MeV},
\label{eq:2p4}
\end{eqnarray}
in agreement with Migdal's original calculation~\cite{migdal1945quadrupole},
where $\kappa$ is the dipole polarizability parameter that accounts for deviations ($\kappa \neq 1$) of the hydrodynamic model from the actual {\sc GDR} effects.
Calculations can be benchmarked with available photo-absorption cross-section data~\cite{exfor,ENDF}.

In this work, we further explore how the {\sc E1} polarizability evolves
from the ground state to the first excitation of selected {\it p}- and {\it sd}-shell nuclides.
Where possible, we perform  $1\hbar\omega$ {\sc SM} calculations, compare with
available data and explore deviations from the hydrodynamic model~\cite{orce2015new,orce2016reply,ngwetsheni2019continuing,ngwetsheni2019combined,ngwetsheni2019how}.
Similar {\sc SM} calculations of the {\sc E1} polarizability for ground states  have already been published in
Ref.~\cite{orce2023global}.

\begin{figure}[!ht]
\begin{center}
\includegraphics[width=5.5cm,height=8.5cm,angle=-0]{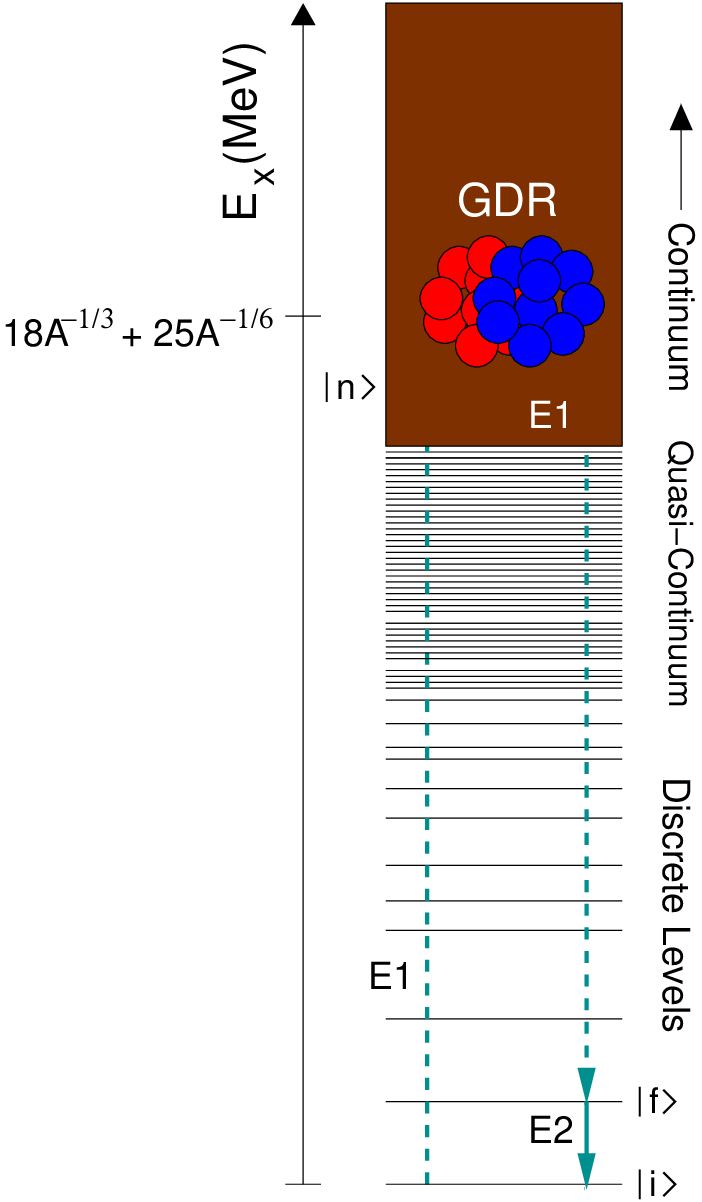}
\caption{Sketch of {\sc E1} virtual excitations around the {\sc GDR} polarizing the final state $\lvert~ f ~\rangle$. The energy of the {\sc GDR} built on excited states is parametrized as $18 A^{-1/3} + 25 A^{-1/6}$ MeV~\cite{gaardhoje1992nuclear}.~\label{fig:GDR}}
\end{center}
\end{figure}

\section{Shell-model Calculations}

Firstly, we deduce new polarization equations for excited states on the same footing as ground states. Applying second-order non-degenerate perturbation theory to the Coulomb-excitation process shown in Fig.~\ref{fig:GDR}, the second-order transition amplitude $b_{i\rightarrow f}^{^{(2)}}$ from  $\lvert i\rangle$ --- again, the ground state --- to
a final excited state $\lvert f\rangle$ is given by~\cite{deBoer1968}
\begin{eqnarray}
b_{_{i\rightarrow f}}^{^{(2)}} = b^{^{(1)}}_{_{i\rightarrow f}} + \sum_{n} b_{_{inf}},
\label{eq:bif2}
\end{eqnarray}
where $b^{^{(1)}}_{_{i\rightarrow f}}$ is proportional to the $\langle i\parallel\hat{E2}\parallel f\rangle$ transitional matrix element and
$\sum\limits_{n} b_{_{inf}}$ is the sum over intermediate states of the interference term between first-order and second-order transitions,
which is proportional to
\begin{eqnarray}
\mathcal{S}(E1) = \frac{1}{2J_i+1}\sum_{J_n, \Delta T}  W ~\frac{\langle i \parallel
\hat{E1} \parallel n\rangle \langle n\parallel \hat{E1} \parallel
f\rangle}{E_n - E_i}&,
\label{eq:se1}
\end{eqnarray}
 where $W=W(\lambda_{in} ~\lambda_{nf} ~J_i ~J_f, \lambda ~J_n)$ \nolinebreak are the corresponding Racah $W$-coefficients~\cite{racah1942theory} ---
with $\lambda_{in} = \lambda_{nf} =1$ for {\sc E1} multipolarity and $\lambda = \lambda_{in} + \lambda_{nf}$\footnote{Racah (and Wigner) coefficients are zero if the triangle inequalities $J_i + J_f \geq J_n$ and $\lambda_{in} + \lambda_{nf} \geq \lambda$ are not satisfied.}--- and the sum extends over the intermediate $J_n$ states connecting both  $\lvert i\rangle$ and $\lvert f\rangle$ states with isovector $E1$ transitions following the general isospin selection rule for electromagnetic transitions: $\Delta T=0, \pm1$~\cite{warburton1969role}. Particularly, for self-conjugate $N=Z$ nuclei the isovector contribution
arises only from $\Delta T=1$ transitions\footnote{Again, because of a vanishing $3j$ symbol for $\Delta T =0$.}, while both $\Delta T =0$ and $\Delta T =1$  have to be considered otherwise.

Furthermore, $\mathcal{S}(E1)$ is connected to  Eq.~\ref{eq:polar2} through
the  reference parameter $\eta_{_{0}}$~\cite{hausser1973e1,eichler1964second,deBoer1968,macdonald1964coulomb,nebel1967second,douglas1967coulomb},
\begin{eqnarray}
\eta_{_{0}} = \sqrt{2/5} ~\dfrac{\sum\limits_{n} \dfrac{\langle i\parallel\hat{E1}\parallel n\rangle \langle n\parallel\hat{E1}\parallel f\rangle}{E_n - E_i}}
{{\sum\limits_{n} \dfrac{|\langle i\parallel\hat{E1}\parallel n\rangle|^2}{E_n - E_i}}},
\label{eq:sigmase1}
\end{eqnarray}
where the relation
\begin{eqnarray}
\eta_{_{0}}= \frac{4\sqrt{\pi}}{3} \frac{\langle i\parallel \hat{E2} \parallel
f\rangle }{ZeR^2}
\label{eq:eta}
\end{eqnarray}
is defined by association with the excitation amplitude $\sum\limits_{n} b_{_{inf}}$~\cite{deBoer1968}.
For simplicity, Eichler originally assumed the closure approximation~\cite{eichler1964second} ---  i.e. $\eta_{_{0}}=1$ for closed-shell nuclei ---
but smaller $\eta_{_{0}}<1$ values are expected because of the random phase of the off-diagonal matrix elements in the numerator of Eq.~\ref{eq:sigmase1}.
For instance, $\eta_{_{0}} \simeq  1/12$ is determined assuming a 2$^+_1$ rotational state and the characteristic energy relation of the {\sc GDR} double peak  for a strongly-deformed prolate nucleus~\cite{macdonald1964coulomb}, reaching  smaller $\eta_{_{0}} \leq  1/12$ values when considering the triaxial degree of freedom. A slightly larger value of $\eta_{_{0}} \approx 0.3$
is determined using the dynamic collective model for a strongly-deformed nucleus~\cite{nebel1967second}.
Generally, the dynamic collective model by Danos and Greiner~\cite{danos1964dynamic}, later extended to spherical nuclei by Weber~\cite{weber1966theorie,huber1967collective}, can be used
to calculate $\eta_{_{0}}$ values, which allows for rotations, surface quadrupole vibrations and higher-energy giant resonance oscillations.
As stated by de Boer and Eichler, \emph{Eq.~\ref{eq:eta} may also provide a useful estimate for the more general case, since the polarization is essentially a nuclear-size effect}.
Moreover, the electric quadrupole   $\langle i\parallel \hat{E2} \parallel f\rangle $ matrix element in Eq.~\ref{eq:eta} naturally arises
from the interference between first-order (E2) and second-order (E1) transitions~\cite{deBoer1968}.

Now using Eqs. \ref{eq:polar2}, \ref{eq:se1}, \ref{eq:sigmase1} and \ref{eq:eta}, new relations for $\alpha_{_{E1}}$ and $\sigma_{_{-2}}$ values can be deduced for
excited states with arbitrary $J$,
\begin{eqnarray}
\alpha_{_{E1}}
&=& 1.11 ~\frac{Z A^{2/3}}{ \langle i\parallel \hat{E2} \parallel f\rangle} ~\sum_{J_n, \Delta T}\frac{\mathcal{S}(E1)}{W(\lambda_{in} ~\lambda_{nf} ~J_i ~J_f, \lambda ~J_n)}, \label{eq:alphanew}\\
\sigma_{_{-2}}
&=& 0.155 ~\frac{Z A^{2/3}}{ \langle i\parallel \hat{E2} \parallel f\rangle}~ \sum_{J_n, \Delta T}\frac{\mathcal{S}(E1)}{W(\lambda_{in} ~\lambda_{nf}~ J_i ~J_f, \lambda ~J_n)}, \label{eq:sigma2new}
\end{eqnarray}
in units of fm$^3$ and  fm$^2$/MeV, respectively.
For instance,  assuming a two-step processes of the type $0^+_{_1} \dashrightarrow  1^-_{_{GDR}} \dashrightarrow  2^+_{_1}$ in even-even nuclei, $W(1 ~1 ~0 ~2, ~2 ~1)=\sqrt{1/3}\sqrt{1/5}$, yielding
\begin{equation}
\sigma_{_{-2}} = 0.6 ~\frac{Z A^{2/3} \mathcal{S}(E1)}{ \langle 0^+_{_1} \parallel \hat{E2} \parallel 2^+_{_1} \rangle}.
\end{equation}

Further, the polarizability parameter $\kappa$ can also be determined in terms of {\sc E1} and {\sc E2} matrix elements, as originally done by H\"ausser and collaborators~\cite{hausser1973e1},
\begin{eqnarray}
\kappa &=& \frac{ \sigma_{_{-2}}}{2.38 \times 10^{-4}~A^{5/3}}  \\
&=& \frac{\sum\limits_{J_n,  \Delta T}\frac{\mathcal{S}(E1)}{W(1,1,J_i,J_f, 2,J_n)}}{0.0015 \frac{A}{Z} ~\langle i\parallel \hat{E2} \parallel f\rangle}.
\label{eq:kappa}
\end{eqnarray}
For the usual case of a final 2$^+_{_1}$ state following $0^+_{_1} \dashrightarrow  1^-_{_{GDR}} \dashrightarrow  2^+_{_1}$,
\begin{equation}
\kappa= \frac{1}{0.00039 ~\frac{A}{Z}}\frac{\mathcal{S}(E1)}{~\langle 0^+_{_1} \parallel \hat{E2} \parallel 2^+_{_1} \rangle}\footnote{This is an equivalent relation to the one deduced by H\"ausser~\cite{hausser1973e1,hausser1974} and Barker~\cite{barger1982gdr,barker1982gdr17O, barker1982gdr18O}, but using $\sigma_{_{-2}}=2.38 \times 10^{-4}~\kappa ~A^{5/3}$ fm$^2$/MeV instead of Levinger's  $\sigma_{_{-2}}=3.5 \times 10^{-4}~\kappa ~A^{5/3}$ fm$^2$/MeV~\cite{levinger1957migdal}, which yields a factor of 0.00058 in the denominator~\cite{barker1982gdr17O}.}.
\end{equation}
Additional Racah coefficients of
\begin{eqnarray}
W(1~1~1~3,~2~2)&=&1/5, ~\label{eq:6Li} \\
W(1~1~3/2~1/2,2~1/2)&=&15\sqrt{1/3}\sqrt{1/5}\sqrt{1/6}\sqrt{1/30}, ~\label{eq:7Li1}\\
W(1~1~3/2~1/2,2~3/2)&=&15/2\sqrt{1/3}\sqrt{1/5}\sqrt{1/15}\sqrt{1/30}, ~\label{eq:7Li2} \\
W(1~1~~5/2~1/2,2~3/2)&=&3/2 \sqrt{1/3}\sqrt{1/15}, ~\label{eq:17O}
\end{eqnarray}
were used for $^6$Li ($1^+_{_1} \dashrightarrow  2^-_{_{GDR}} \dashrightarrow  3^+_{_1}$), $^7$Li ($3/2^-_{_1} \dashrightarrow  1/2^+_{_{GDR}} \dashrightarrow  1/2^-_{_1}$
and $3/2^-_{_1} \dashrightarrow  3/2^+_{_{GDR}} \dashrightarrow  1/2^-_{_1}$ ) and $^{17}$O ($5/2^+_{_1} \dashrightarrow  3/2^-_{_{GDR}} \dashrightarrow  1/2^+_{_1}$ ).

Accordingly, Eqs. \ref{eq:alphanew}, ~\ref{eq:sigma2new} and \ref{eq:kappa} allow the general calculation of  $\alpha_{_{E1}}$,  $\sigma_{_{-2}}$ and $\kappa$ values for excited states
using {\sc E1} and {\sc E2} matrix elements computed by various theoretical models, comparison with sum rules and Coulomb-excitation measurements.
Here, accurate Coulomb-excitation measurements of  second-order contributions to the inelastic cross sections could be used to benchmark the polarizability of excited states~\cite{eichler1964second}.

\begin{figure*}[!ht]
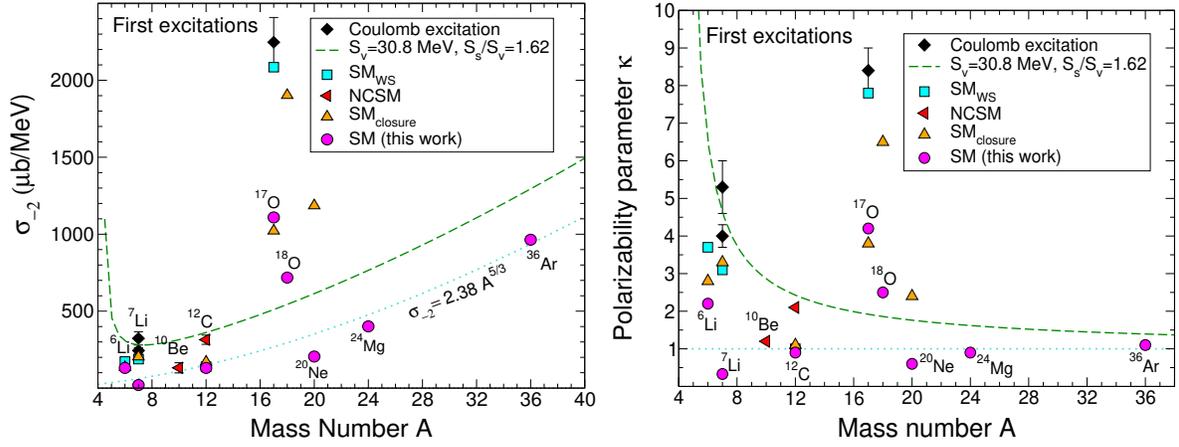

\begin{center}
\includegraphics[width=7.8cm,height=5.8cm,angle=-0]{sigma2_1stexcitation_SM_exp_Nov23.eps}
\hspace{0.cm}
\includegraphics[width=7.5cm,height=5.8cm,angle=-0]{kappa_exc_Nov23.eps}
\caption{Experimental and calculated $\sigma_{_{-2}}$ (left) and $\kappa$ (right) values for the first excitation of selected $p$ and {\it sd-}shell nuclei.
The experimental points (diamonds) are taken from Coulomb-excitation data~\cite{hausser1973e1,vermeer1984coulomb,vermeer1984experimental,kuehner1982measurement}.
The {\sc SM} calculations consider the closure approximation (up triangles)~\cite{barger1982gdr,barker1982gdr17O,barker1982gdr18O},
Woods-Saxon wave functions (squares)~\cite{barke1989investigation},
 {\sc NCSM} (left triangles)~\cite{raju2018reorientation,orce2012reorientation} and current work (circles). For comparison, the hydrodynamic-model prediction for $\kappa=1$ in Eq.~\ref{eq:2p4} is shown by dotted lines together with the leptodermous trend provided by the finite-range droplet model ({\sc FRDM}) with symmetry energy coefficients $S_v= 30.8$ MeV and $S_s/S_v = 1.62$~\cite{moller1995atomic} (dashed lines).}
\label{fig:expsm}
\end{center}
\end{figure*}

\begin{table*}[!ht]
\begin{center}
\caption{Calculated $\sigma_{_{-2}}^{^{SM}}$ (column 5) and $\kappa^{^{SM}}$ (column 10) values of first-excited states in selected $p-sd$ shell nuclei together with their
corresponding $\mathcal{S}(E1)^{^{SM}}$ and $\langle i\parallel\hat{E2} \parallel f\rangle^{^{SM}}$  (columns 6 and 7).
Experimental $\langle i\parallel\hat{E2}\parallel f\rangle^{^{exp}}$~\cite{sonzogni2007nndc}  and $\kappa^{^{exp}}$ values (columns 8 an 9, respectively) as well as
previous {\sc SM} calculations (column 11) are listed for comparison. \label{tab:tab1}}
\begingroup
\setlength\textwidth\textheight
\setlength{\tabcolsep}{1.pt}
\colorbox{gray!70!yellow!10}
{\footnotesize
\begin{tabular}{|ccccccccccc|}
\hline \hline
Nucleus & $J^{\pi}_i$  &  $J^{\pi}_f$  & $E_{ex}$ &   $\sigma_{_{-2}}^{^{SM}}$ &  $\mathcal{S}(E1)^{^{SM}}$  &
$\langle i\parallel\hat{E2}\parallel f\rangle^{^{SM}}$ & $\langle i\parallel\hat{E2}\parallel f\rangle^{^{exp}}$ &
$\kappa^{exp}$ & $\kappa^{^{SM}}$  &  $\kappa^{previous ~SM}$          \\
        &      &      & MeV         &   $\mu$b/MeV               &     e$^2$fm$^2$/MeV   &    e$^2$fm$^2$/MeV &                     e~fm$^2$ & &         &  \\
\hline
${^6}$Li    &  1$^+_{_1}$     & 3$^+$     & 2.19  &       131      &   0.0135    & 7.88        &   8.65(48)       &  --    &  2.2  &    1.3, 3.7~\cite{barke1989investigation}, 2.8~\cite{barger1982gdr} \\
\hline
${^7}$Li    &  3/2$_{_1}^{-}$ & 1/2$_{_1}^{-}$ & 0.48  &   20  &  0.0007  &  6.57 & 5.60(56) &  4.0(3)~\cite{vermeer1984coulomb,vermeer1984experimental}, 5.3(7)~\cite{hausser1973e1}                                              &                          0.3     &        1.5, 3.1~\cite{barke1989investigation}, 3.3~\cite{barger1982gdr}    \\
\hline
${^{12}}$C  &  0$_{_1}^{+}$  & 2$_{_1}^{+}$   & 4.44  &   131   & 0.0045  & 6.52 & 6.30(16) &   --                              & 0.9  & 0.9, 1.0~\cite{barke1989investigation}, 1.1~\cite{barger1982gdr}, 2.1(2)*~\cite{raju2018reorientation}               \\
\hline
${^{17}}$O  & 5/2$_{_1}^{+}$ & 1/2$_{_1}^{+}$ & 0.87  &    1109  &  0.0103  & 3.40 & 3.55(3)  & 8.4(6)~\cite{kuehner1982measurement}   & 4.2  & 3.8~\cite{barker1982gdr17O}, 2.3, 5.7, 7.8**~\cite{barke1989investigation}                 \\
\hline
${^{18}}$O  &  0$_{_1}^{+}$  & 2$_{_1}^{+}$   & 1.98 &  718  & 0.0088 &  4.05  &  6.56(29) &  --                                & 2.5  &  6.5~\cite{barker1982gdr18O}   \\
\hline
${^{20}}$Ne &  0$_{_1}^{+}$  & 2$_{_1}^{+}$   & 1.63 & 205  & $-$0.0089  & $-$19.25  &  18.25(44)   &  --  &  0.6  &  3.4 \cite{barker1982gdr18O}  \\
\hline
${^{24}}$Mg &  0$_{_1}^{+}$  & 2$_{_1}^{+}$   & 1.37 &  401  & $-$0.0146  & $-$21.80   &  20.91(22)   &  --  &  0.9  &   --    \\
\hline
${^{36}}$Ar & 0$_{_1}^{+}$  & 2$_{_1}^{+}$   & 1.97 & 964  & $-$0.0151  & $-$18.48  & 17.35(46)  &        --                        & 1.1  &   --    \\
\hline \hline
\end{tabular}
}\endgroup
 \begin{tablenotes}
      \footnotesize
      \item * The quoted value of $\kappa(2^+_{_1})=2.1(2)$ in Ref.~\cite{raju2018reorientation} was determined by {\sc  NCSM}
      calculations using
      the $NN + 3N350-srg2.0$ $\chi EFT$ interaction with a basis size of $N_{max}=4$. A similar  value of $\kappa(2^+_{_1})=2.2(2)$
      is also provided in Ref.~\cite{raju2018reorientation} using the $NN~N^4NLO500-srg2.4$ interaction.
      \item ** Only result that includes configuration mixing in Refs.~\cite{barger1982gdr,barker1982gdr17O,barker1982gdr18O,barke1989investigation}.
      \item  All quoted  $\sigma_{_{-2}}$  and $\kappa$ values presented in this work are related to Eq.~\ref{eq:2p4}.
    \end{tablenotes}
\end{center}
\end{table*}

Shell-model calculations of  $\sigma_{_{-2}}$ and $\kappa$ values for first low-lying excitations were previously performed
in light nuclei~\cite{barger1982gdr,barker1982gdr17O,barker1982gdr18O,barke1989investigation,raju2018reorientation,orce2012reorientation} ---
as shown in Fig.~\ref{fig:expsm} --- and used in the analysis of Coulomb-excitation studies in order to treat adequately the {\sc GDR} effect~\cite{hausser1973e1,vermeer1984coulomb,vermeer1984experimental,kuehner1982measurement,raju2018reorientation,hausser1972nuclear,orce2012reorientation,disdier1971projectile,vermeer1982giant,vermeer1983electric}.  Large $\kappa>1$ values were predicted for the $J=1/2_1$ excited state in  $^7$Li and $^{17}$O~\cite{barger1982gdr,barker1982gdr17O,barke1989investigation}, in agreement with Coulomb-excitation measurements~\cite{hausser1973e1,vermeer1984coulomb,vermeer1984experimental,kuehner1982measurement,hausser1972nuclear}.
In more detail, the first {\sc SM} calculations in $0sd$-shell nuclei~\cite{barger1982gdr} performed for $^{17}$O~\cite{barker1982gdr17O} and $^{18}$O~\cite{barker1982gdr18O} assumed the closure approximation (up triangles in Fig.~\ref{fig:expsm}), in which all the {\sc E1} strength is concentrated at the {\sc GDR} energy.
A more exhaustive approach  considered a realistic Hamiltonian with harmonic-oscillator and Woods-Saxon (squares in Fig.~\ref{fig:expsm}) single-particle
wave functions restricted to the lowest configuration~\cite{barke1989investigation}, which included configuration mixing
only for the simpler case of $^{17}$O and presented a  better correlation with experimental values.

Recently, no-core shell model ({\sc NCSM}) calculations  of $\kappa$ values for excited states in $^{10}$Be~\cite{orce2012reorientation} and $^{12}$C~\cite{raju2018reorientation} (left triangles in Fig.~\ref{fig:expsm}) were performed using $2N$ and $2N + 3N$ forces~\cite{navratil2007local,roth2014evolved,entem2003accurate,entem2017high,entem2015peripheral,bogner2007similarity}
with $N_{max}=4$ basis sizes for natural and
$N_{max}=5$ for  unnatural parity states. These {\it ab initio} calculations included {\sc E1} matrix elements connecting about 30 1$^-$ states up to 30 MeV.
More advanced {\it ab initio} calculations involve the Lanczos-continued-fraction algorithm~\cite{haydock1974inverse,marchisio2003efficient} applied to Eq.~\ref{eq:se1} by Navr\'atil,
which sum up contributions of all the excited states~\cite{cebotobe}.
For the ground states of $^{9}$Be and $^{12}$C, values of $\kappa(g.s.)=3.4(8)$ and $\kappa(g.s.)=1.6(2)$
were predicted, respectively, in agreement with photo-absorption cross-section data~\cite{levinger1957migdal,nathans1953excitation,fuller1985photonuclear}.
Values of  $\kappa(2^+_{_1})=1.2(3)$ and  $\kappa(2_{_1}^+) = 2.1(2)$ were calculated for the
first excited states in $^{10}$Be and $^{12}$C, respectively.\\


In the present work,  full 1$\hbar\omega$ {\sc SM} calculations of the dipole polarizability for the first excitation in selected $p-$ and $sd$-shell nuclei have been performed with the {\sc OXBASH} code~\cite{brown1988computer} using the {\sc WBP}~\cite{warburton1992effective} and  {\sc FSU}~\cite{lubna2019structure,lubna2020evolution,brown2022nuclear} Hamiltonians and the $spsdpf$ model space.
Shell model calculations for $A\leq12$ nuclides arise from the {\sc WBP} interaction whereas  for $A\geq17$ we quote results using the {\sc FSU} interaction.
Indeed, the {\sc FSU} Hamiltonian starts with the {\sc WBP} Hamiltonian, fitting over 270 experimental levels from $^{13}$C to $^{51}$Ti, and additionally includes
particle-hole states originating from cross-shell excitations
that give rise to intruder states.
In essence, both Hamiltonians are the same near $^{16}$O, with the {\sc WBP} interaction presenting --- as for ground states~\cite{orce2023global} ---
slightly smaller polarizability values in the middle and end of the {\it sd} shell.


The general procedure
involves the calculation of all the {\sc E1} and {\sc E2} matrix elements following Eqs.~\ref{eq:sigma2new} and \ref{eq:kappa}.  Similar {\sc SM} calculations were performed for ground states and are explained in detail in Ref.~\cite{orce2023global}. For self-conjugate nuclei, we calculate {\sc E1} matrix elements connecting all
the intermediate {\sc GDR} states with $\Delta T=1$;
for instance,
a total of  6770 $\langle 0^+_{_1} \parallel\hat{E1}\parallel 1^-_{_n}\rangle$ and $\langle 1^-_{_n}\parallel\hat{E1}\parallel 2^+_{_1}\rangle$  matrix elements in $^{36}$Ar.
Similarly, we calculate all possible {\sc E1} matrix elements from the various {\sc GDR} intermediate states for other nuclei, including $\Delta T=0,1$ isovector transitions.
For consistency, all $\langle i\parallel\hat{E2}\parallel f\rangle$ matrix elements required in the $0\hbar\omega$ calculation of
$\sigma_{_{-2}}^{^{SM}}$  and $\kappa^{^{SM}}$  values are also computed using {\sc OXBASH} with isoscalar {\sc E2} effective charges of $e^p_{_{eff}}+e^n_{_{eff}}=1.55$ and $e^p_{_{eff}}+e^n_{_{eff}}=1.81$ for the $p$~\cite{bassel19820+} and $sd$~\cite{richter2008sd} shells, respectively.

Results arising from the {\sc WBP} interaction have been benchmarked for compatibility with available {\sc NCSM} calculations in
$^{12}$C~\cite{raju2018reorientation} using the {\sc $NN + 3N350-srg2.0$}~\cite{navratil2007local,roth2014evolved,entem2003accurate,bogner2007similarity} and {\sc $NN~N^4NLO500-srg2.4$} ~\cite{entem2017high,entem2015peripheral,bogner2007similarity} ~$\chi EFT$ interactions,
yielding a difference of $\approx27\%$ for the product of {\sc E1} matrix elements in Eq.~\ref{eq:se1}.
Nevertheless, the highest 1$^-$ state calculated with {\sc OXBASH} is at $\approx 65$ MeV,
whereas for the {\sc NCSM} it corresponds to $\approx30$ MeV. Below 30 MeV, the sum of {\sc E1} strengths reach converging values of
$\approx1.3$ e$^2$fm$^2$ and $\approx0.16$ e$^2$fm$^2$ for $0^+_{_1} \dashrightarrow 1^-_{_n}$ and $1^-_{_n} \dashrightarrow 2^+_{_1}$ transitions, respectively, with $\mathcal{S}(E1)$ differing only by $\approx8\%$.\\

Further, {\sc SM} calculations generally underestimate collective {\sc E2} properties~\cite{barke1989investigation,epelbaum2012structure,stroberg2022systematics,henderson2022coulomb,sarma2023ab},
which may result in artificially large $\kappa$ values. Therefore, the calculated $\langle i \parallel \hat{E2} \parallel f \rangle^{^{SM}}$ matrix elements should also be compared with {\sc NCSM} calculations and data. Considering the  $^{12}$C testing ground, {\sc NCSM} calculations with the {\sc $NN+3N350-srg2.0$} interaction at $N_{max}=4$ underestimates by 23\%, $\langle 0^+_{_1}\parallel \hat{E2} \parallel 2^+_{_1}\rangle^{^{NCSM}} = 0.04786$ eb, the well-known
 $\langle 0^+_{_1}\parallel \hat{E2} \parallel 2^+_{_1}\rangle^{^{exp}}=0.06247(63)$ eb~\cite{pritychenko2016tables},
whereas {\sc OXBASH} calculations with the {\sc WBP} interaction slightly overestimates it by 4\%, $\langle 0^+_{_1}\parallel \hat{E2} \parallel 2^+_{_1}\rangle^{^{WBP}} = 0.06522$ eb.
This discrepancy encourages further {\sc NCSM} calculations using a new generation of $\chi EFT$ interactions, the Lanczos-continued-fraction algorithm and reaching higher excitation energies and larger $N_{max}$ basis sizes~\cite{sarma2023ab}.
As shown in Table~\ref{tab:tab1}, experimental $\langle i\parallel \hat{E2} \parallel f\rangle^{^{exp}}$ and calculated $\langle i\parallel \hat{E2} \parallel f\rangle^{^{SM}}$ matrix elements are in overall agreement  with the exception of $^{18}$O, where mixing with the intruder states is not included~\cite{brown2017oxygen,lawson1976structure,ellis1970weak,engeland1972weak}.
Table~\ref{tab:tab1} also lists
$\sigma_{_{-2}}^{^{SM}}$ and $\kappa^{^{SM}}$ values for the first excitation of selected $p$ and {\it sd-}shell nuclei with their corresponding $\mathcal{S}(E1)^{^{SM}}$ value.
For comparison, theoretical results are also shown in Fig.~\ref{fig:expsm} together with available measurements of $\sigma_{_{-2}}^{^{exp}}$ (left) and $\kappa^{^{exp}}$ (right) values.



\section{Discussion and Conclusions}

Overall, $\sigma_{_{-2}}^{^{SM}}$ and $\kappa^{^{SM}}$ values calculated for the first excited states in selected light nuclei align with the smooth, global trend predicted by Eq.~\Ref{eq:2p4} (dotted $\kappa=1$ lines in Fig.~\ref{fig:expsm}), in agreement with Migdal's original calculation arising from the combination of the hydrodynamic model and second-order non-degenerate perturbation theory~\cite{migdal1945quadrupole};
hence, validating modern Coulomb-excitation codes~\cite{cline2012gosia} used to extract collective properties of $p$- and $sd$-shell nuclei~\cite{spear1981static}.

Deviations from simple hydrodynamic-model estimates are, nonetheless, observed at the beginning of the $p$- and $sd$-shells, where anomalously large $\sigma_{_{-2}}$ and $\kappa$ values are calculated for the excited states of $^{6,7}$Li and $^{17,18}$O. For the odd-mass nuclei, this may be associated with the slightly unbound particle~\cite{barke1989investigation,barker1984decay}  --- whose wave function  extends far apart from the $\alpha$-cluster configurations, i.e. $\alpha+d$, $\alpha+t$,  $4\alpha+n$ in $^{6}$Li, $^{7}$Li and $^{17}$O, respectively --- as inferred from the dipole resonances observed at relatively low excitation energies~\cite{nakayama2001dipole,burda2010resonance}.
Such cluster structures were avoided in the 1$\hbar\omega$ cross-shell fits to the {\sc WBP} and {\sc FSU} interactions because of potential distortion in the  $A=5-9$ region.
These deviations from the {\sc GDR} effect are not surprising considering
the fragmentation of the {\sc GDR} spectrum into different 1p-1h states~\cite{eramzhyan1986giant}, which include the possibility of
$\alpha$-cluster configurations~\cite{neudatchin1979clustering,he2014giant}  and the virtual breakup into the continuum~\cite{smilansky1972role,weller1985electromagnetic};
the latter supporting the breakup into the $\alpha$-t continuum as the the main contribution to the polarizability in $^7$Li.

The reasons behind the large $\sigma_{_{-2}}^{^{SM}}$ and $\kappa^{^{SM}}$ values computed in $^{17,18}$O deserve further investigation.
For $^{18}$O, this is probably related to the aforementioned mixing of intruder states lacking in the $sd$-shell wave functions.
However, the anomalously large $\kappa$ value measured for the first excitation of $^{17}$O~\cite{kuehner1982measurement} suggests an alternative physical origin.
As suggested by  Kuehner {\it et al.}~\cite{kuehner1982measurement} and Ball {\it et al.}~\cite{ball1982dbla}, a value slightly larger than $\kappa=1$ attributed to the
first 2$^+_1$ excitation at 1.982 MeV in $^{18}$O could explain the long-standing $\approx10$\% discrepancy between the smaller
$B(E2; 0^+_1 \rightarrow 2^+_1) = 0.00421(9)$ e$^2$b$^2$ determined from seven Coulomb-excitation
measurements~\cite{raman1987transition} and the larger one, 0.00476(11) e$^2$b$^2$, extracted from a high-precision lifetime measurement~\cite{ball1982dbla}.
The overall lack of high-sensitive measurements of the nuclear polarizability of excited states prevent further conclusions.

In conclusion, we present novel equations of $\sigma_{_{-2}}$ and $\alpha_{_{E1}}$ values for excited states on equal footing to ground states by calculating {\sc E1} and {\sc E2} matrix elements.
We apply this framework and perform full 1$\hbar\omega$ {\sc SM} calculations of $\sigma_{_{-2}}^{^{SM}}$ and $\kappa^{^{SM}}$ values up to $^{36}$Ar, with reasonable agreement with previous
{\sc SM} calculations done up to $^{20}$Ne. Except the anomalous cases of $^{17,18}$O, results follow the global trend predicted by Migdal~\cite{migdal1945quadrupole}. Dedicated Coulomb-excitation measurements with increasing sensitivity are relevant in order to elucidate the reasons behind large dipole polarizabilities that may affect quadrupole collective properties~\cite{orce2020polarizability,alder1975north,orce2021reorientation}. These measurements could be done with the new {\sc GAMKA} array at iThemba LABS in South Africa and elsewhere.
More detailed \emph{ab initio} calculations with uncertainty quantification above $^{12}$C --- the heaviest nucleus where the dipole polarizability for excited states has been computed from first principles --- are needed to benchmark our results.\\

\section*{Acknowledgements}

We acknowledge B. A. Brown for physics discussion and supplying the {\sc OXBASH} code  to determine {\sc E1} and {\sc E2} matrix elements.
We also thank the handling Editors and all the anonymous Referees of the Journal of Physics G for their constructive comments and suggestions. \\

\bibliographystyle{apsrev4-2}
\bibliography{jpgexc}

\begin{thebibliography}{120}%
\makeatletter
\providecommand \@ifxundefined [1]{%
 \@ifx{#1\undefined}
}%
\providecommand \@ifnum [1]{%
 \ifnum #1\expandafter \@firstoftwo
 \else \expandafter \@secondoftwo
 \fi
}%
\providecommand \@ifx [1]{%
 \ifx #1\expandafter \@firstoftwo
 \else \expandafter \@secondoftwo
 \fi
}%
\providecommand \natexlab [1]{#1}%
\providecommand \enquote  [1]{``#1''}%
\providecommand \bibnamefont  [1]{#1}%
\providecommand \bibfnamefont [1]{#1}%
\providecommand \citenamefont [1]{#1}%
\providecommand \href@noop [0]{\@secondoftwo}%
\providecommand \href [0]{\begingroup \@sanitize@url \@href}%
\providecommand \@href[1]{\@@startlink{#1}\@@href}%
\providecommand \@@href[1]{\endgroup#1\@@endlink}%
\providecommand \@sanitize@url [0]{\catcode `\\12\catcode `\$12\catcode
  `\&12\catcode `\#12\catcode `\^12\catcode `\_12\catcode `\%12\relax}%
\providecommand \@@startlink[1]{}%
\providecommand \@@endlink[0]{}%
\providecommand \url  [0]{\begingroup\@sanitize@url \@url }%
\providecommand \@url [1]{\endgroup\@href {#1}{\urlprefix }}%
\providecommand \urlprefix  [0]{URL }%
\providecommand \Eprint [0]{\href }%
\providecommand \doibase [0]{https://doi.org/}%
\providecommand \selectlanguage [0]{\@gobble}%
\providecommand \bibinfo  [0]{\@secondoftwo}%
\providecommand \bibfield  [0]{\@secondoftwo}%
\providecommand \translation [1]{[#1]}%
\providecommand \BibitemOpen [0]{}%
\providecommand \bibitemStop [0]{}%
\providecommand \bibitemNoStop [0]{.\EOS\space}%
\providecommand \EOS [0]{\spacefactor3000\relax}%
\providecommand \BibitemShut  [1]{\csname bibitem#1\endcsname}%
\let\auto@bib@innerbib\@empty
\bibitem [{\citenamefont {Dietrich}\ and\ \citenamefont
  {Berman}(1988)}]{dietrich1988atlas}%
  \BibitemOpen
  \bibfield  {author} {\bibinfo {author} {\bibfnamefont {S.~S.}\ \bibnamefont
  {Dietrich}}\ and\ \bibinfo {author} {\bibfnamefont {B.~L.}\ \bibnamefont
  {Berman}},\ }\href@noop {} {\bibfield  {journal} {\bibinfo  {journal} {Atomic
  Data and Nuclear Data Tables}\ }\textbf {\bibinfo {volume} {38}},\ \bibinfo
  {pages} {199} (\bibinfo {year} {1988})}\BibitemShut {NoStop}%
\bibitem [{\citenamefont {Plujko}\ \emph {et~al.}(2018)\citenamefont {Plujko},
  \citenamefont {Gorbachenko}, \citenamefont {Capote},\ and\ \citenamefont
  {Dimitriou}}]{plujko2018giant}%
  \BibitemOpen
  \bibfield  {author} {\bibinfo {author} {\bibfnamefont {V.~A.}\ \bibnamefont
  {Plujko}}, \bibinfo {author} {\bibfnamefont {O.~M.}\ \bibnamefont
  {Gorbachenko}}, \bibinfo {author} {\bibfnamefont {R.}~\bibnamefont
  {Capote}},\ and\ \bibinfo {author} {\bibfnamefont {P.}~\bibnamefont
  {Dimitriou}},\ }\href@noop {} {\bibfield  {journal} {\bibinfo  {journal}
  {Atomic Data and Nuclear Data Tables}\ }\textbf {\bibinfo {volume} {123}},\
  \bibinfo {pages} {1} (\bibinfo {year} {2018})}\BibitemShut {NoStop}%
\bibitem [{\citenamefont {Kawano}\ \emph {et~al.}(2020)\citenamefont {Kawano},
  \citenamefont {Cho}, \citenamefont {Dimitriou}, \citenamefont {Filipescu},
  \citenamefont {Iwamoto}, \citenamefont {Plujko}, \citenamefont {Tao},
  \citenamefont {Utsunomiya}, \citenamefont {Varlamov}, \citenamefont {Xu}
  \emph {et~al.}}]{kawano2020iaea}%
  \BibitemOpen
  \bibfield  {author} {\bibinfo {author} {\bibfnamefont {T.}~\bibnamefont
  {Kawano}}, \bibinfo {author} {\bibfnamefont {Y.}~\bibnamefont {Cho}},
  \bibinfo {author} {\bibfnamefont {P.}~\bibnamefont {Dimitriou}}, \bibinfo
  {author} {\bibfnamefont {D.}~\bibnamefont {Filipescu}}, \bibinfo {author}
  {\bibfnamefont {N.}~\bibnamefont {Iwamoto}}, \bibinfo {author} {\bibfnamefont
  {V.}~\bibnamefont {Plujko}}, \bibinfo {author} {\bibfnamefont
  {X.}~\bibnamefont {Tao}}, \bibinfo {author} {\bibfnamefont {H.}~\bibnamefont
  {Utsunomiya}}, \bibinfo {author} {\bibfnamefont {V.}~\bibnamefont
  {Varlamov}}, \bibinfo {author} {\bibfnamefont {R.}~\bibnamefont {Xu}}, \emph
  {et~al.},\ }\href@noop {} {\bibfield  {journal} {\bibinfo  {journal} {Nuclear
  Data Sheets}\ }\textbf {\bibinfo {volume} {163}},\ \bibinfo {pages} {109}
  (\bibinfo {year} {2020})}\BibitemShut {NoStop}%
\bibitem [{\citenamefont {Ishkhanov}\ and\ \citenamefont
  {Kapitonov}(2021)}]{ishkhanov2021giant}%
  \BibitemOpen
  \bibfield  {author} {\bibinfo {author} {\bibfnamefont {B.~S.}\ \bibnamefont
  {Ishkhanov}}\ and\ \bibinfo {author} {\bibfnamefont {I.~M.}\ \bibnamefont
  {Kapitonov}},\ }\href@noop {} {\bibfield  {journal} {\bibinfo  {journal}
  {Physics-Uspekhi}\ }\textbf {\bibinfo {volume} {64}},\ \bibinfo {pages} {141}
  (\bibinfo {year} {2021})}\BibitemShut {NoStop}%
\bibitem [{\citenamefont {Migdal}(1945)}]{migdal1945quadrupole}%
  \BibitemOpen
  \bibfield  {author} {\bibinfo {author} {\bibfnamefont {A.}~\bibnamefont
  {Migdal}},\ }\href@noop {} {\bibfield  {journal} {\bibinfo  {journal}
  {Zhurnal Eksperimentalnoi i Teoreticheskoi Fiziki}\ }\textbf {\bibinfo
  {volume} {15}},\ \bibinfo {pages} {81} (\bibinfo {year} {1945})}\BibitemShut
  {NoStop}%
\bibitem [{\citenamefont {Goldhaber}\ and\ \citenamefont
  {Teller}(1948)}]{goldhaber1948nuclear}%
  \BibitemOpen
  \bibfield  {author} {\bibinfo {author} {\bibfnamefont {M.}~\bibnamefont
  {Goldhaber}}\ and\ \bibinfo {author} {\bibfnamefont {E.}~\bibnamefont
  {Teller}},\ }\href@noop {} {\bibfield  {journal} {\bibinfo  {journal}
  {Physical Review}\ }\textbf {\bibinfo {volume} {74}},\ \bibinfo {pages}
  {1046} (\bibinfo {year} {1948})}\BibitemShut {NoStop}%
\bibitem [{\citenamefont {Steinwedel}\ \emph {et~al.}(1950)\citenamefont
  {Steinwedel}, \citenamefont {Jensen},\ and\ \citenamefont
  {Jensen}}]{steinwedel1950nuclear}%
  \BibitemOpen
  \bibfield  {author} {\bibinfo {author} {\bibfnamefont {H.}~\bibnamefont
  {Steinwedel}}, \bibinfo {author} {\bibfnamefont {J.~H.~D.}\ \bibnamefont
  {Jensen}},\ and\ \bibinfo {author} {\bibfnamefont {P.}~\bibnamefont
  {Jensen}},\ }\href@noop {} {\bibfield  {journal} {\bibinfo  {journal}
  {Physical Review}\ }\textbf {\bibinfo {volume} {79}},\ \bibinfo {pages}
  {1019} (\bibinfo {year} {1950})}\BibitemShut {NoStop}%
\bibitem [{\citenamefont {Levinger}\ and\ \citenamefont
  {Kent}(1954)}]{levinger1954independent}%
  \BibitemOpen
  \bibfield  {author} {\bibinfo {author} {\bibfnamefont {J.~S.}\ \bibnamefont
  {Levinger}}\ and\ \bibinfo {author} {\bibfnamefont {D.~C.}\ \bibnamefont
  {Kent}},\ }\href@noop {} {\bibfield  {journal} {\bibinfo  {journal} {Physical
  Review}\ }\textbf {\bibinfo {volume} {95}},\ \bibinfo {pages} {418} (\bibinfo
  {year} {1954})}\BibitemShut {NoStop}%
\bibitem [{\citenamefont {Balashov}(1962)}]{balashov1962relation}%
  \BibitemOpen
  \bibfield  {author} {\bibinfo {author} {\bibfnamefont {V.~V.}\ \bibnamefont
  {Balashov}},\ }\href@noop {} {\bibfield  {journal} {\bibinfo  {journal}
  {Zhurnal {\^E}ksperimental'noi i Teoreticheskoi Fiziki}\ }\textbf {\bibinfo
  {volume} {42}},\ \bibinfo {pages} {275} (\bibinfo {year} {1962})}\BibitemShut
  {NoStop}%
\bibitem [{\citenamefont {Danos}\ and\ \citenamefont
  {Fuller}(1965)}]{danos1965photonuclear}%
  \BibitemOpen
  \bibfield  {author} {\bibinfo {author} {\bibfnamefont {M.}~\bibnamefont
  {Danos}}\ and\ \bibinfo {author} {\bibfnamefont {E.~G.}\ \bibnamefont
  {Fuller}},\ }\href@noop {} {\bibfield  {journal} {\bibinfo  {journal} {Annual
  Review of Nuclear Science}\ }\textbf {\bibinfo {volume} {15}},\ \bibinfo
  {pages} {29} (\bibinfo {year} {1965})}\BibitemShut {NoStop}%
\bibitem [{\citenamefont {Orce}(2022)}]{orce2022competition}%
  \BibitemOpen
  \bibfield  {author} {\bibinfo {author} {\bibfnamefont {J.~N.}\ \bibnamefont
  {Orce}},\ }\href@noop {} {\bibfield  {journal} {\bibinfo  {journal} {Atomic
  Data and Nuclear Data Tables}\ }\textbf {\bibinfo {volume} {145}},\ \bibinfo
  {pages} {101511} (\bibinfo {year} {2022})}\BibitemShut {NoStop}%
\bibitem [{\citenamefont {Rossi}\ \emph {et~al.}(2013)\citenamefont {Rossi},
  \citenamefont {Adrich}, \citenamefont {Aksouh}, \citenamefont {Alvarez-Pol},
  \citenamefont {Aumann}, \citenamefont {Benlliure}, \citenamefont
  {B{\"o}hmer}, \citenamefont {Boretzky}, \citenamefont {Casarejos},
  \citenamefont {Chartier} \emph {et~al.}}]{rossi2013measurement}%
  \BibitemOpen
  \bibfield  {author} {\bibinfo {author} {\bibfnamefont {D.~M.}\ \bibnamefont
  {Rossi}}, \bibinfo {author} {\bibfnamefont {P.}~\bibnamefont {Adrich}},
  \bibinfo {author} {\bibfnamefont {F.}~\bibnamefont {Aksouh}}, \bibinfo
  {author} {\bibfnamefont {H.}~\bibnamefont {Alvarez-Pol}}, \bibinfo {author}
  {\bibfnamefont {T.}~\bibnamefont {Aumann}}, \bibinfo {author} {\bibfnamefont
  {J.}~\bibnamefont {Benlliure}}, \bibinfo {author} {\bibfnamefont
  {M.}~\bibnamefont {B{\"o}hmer}}, \bibinfo {author} {\bibfnamefont
  {K.}~\bibnamefont {Boretzky}}, \bibinfo {author} {\bibfnamefont
  {E.}~\bibnamefont {Casarejos}}, \bibinfo {author} {\bibfnamefont
  {M.}~\bibnamefont {Chartier}}, \emph {et~al.},\ }\href@noop {} {\bibfield
  {journal} {\bibinfo  {journal} {Physical Review Letters}\ }\textbf {\bibinfo
  {volume} {111}},\ \bibinfo {pages} {242503} (\bibinfo {year}
  {2013})}\BibitemShut {NoStop}%
\bibitem [{\citenamefont {Tamii}\ \emph {et~al.}(2011)\citenamefont {Tamii},
  \citenamefont {Poltoratska}, \citenamefont {von Neumann-Cosel}, \citenamefont
  {Fujita}, \citenamefont {Adachi}, \citenamefont {Bertulani}, \citenamefont
  {Carter}, \citenamefont {Dozono}, \citenamefont {Fujita}, \citenamefont
  {Fujita} \emph {et~al.}}]{tamii2011complete}%
  \BibitemOpen
  \bibfield  {author} {\bibinfo {author} {\bibfnamefont {A.}~\bibnamefont
  {Tamii}}, \bibinfo {author} {\bibfnamefont {I.}~\bibnamefont {Poltoratska}},
  \bibinfo {author} {\bibfnamefont {P.}~\bibnamefont {von Neumann-Cosel}},
  \bibinfo {author} {\bibfnamefont {Y.}~\bibnamefont {Fujita}}, \bibinfo
  {author} {\bibfnamefont {T.}~\bibnamefont {Adachi}}, \bibinfo {author}
  {\bibfnamefont {C.}~\bibnamefont {Bertulani}}, \bibinfo {author}
  {\bibfnamefont {J.}~\bibnamefont {Carter}}, \bibinfo {author} {\bibfnamefont
  {M.}~\bibnamefont {Dozono}}, \bibinfo {author} {\bibfnamefont
  {H.}~\bibnamefont {Fujita}}, \bibinfo {author} {\bibfnamefont
  {K.}~\bibnamefont {Fujita}}, \emph {et~al.},\ }\href@noop {} {\bibfield
  {journal} {\bibinfo  {journal} {Physical Review Letters}\ }\textbf {\bibinfo
  {volume} {107}},\ \bibinfo {pages} {062502} (\bibinfo {year}
  {2011})}\BibitemShut {NoStop}%
\bibitem [{\citenamefont {Roca-Maza}\ \emph {et~al.}(2015)\citenamefont
  {Roca-Maza}, \citenamefont {Vi{\~n}as}, \citenamefont {Centelles},
  \citenamefont {Agrawal}, \citenamefont {Colo}, \citenamefont {Paar},
  \citenamefont {Piekarewicz},\ and\ \citenamefont
  {Vretenar}}]{roca2015neutron}%
  \BibitemOpen
  \bibfield  {author} {\bibinfo {author} {\bibfnamefont {X.}~\bibnamefont
  {Roca-Maza}}, \bibinfo {author} {\bibfnamefont {X.}~\bibnamefont
  {Vi{\~n}as}}, \bibinfo {author} {\bibfnamefont {M.}~\bibnamefont
  {Centelles}}, \bibinfo {author} {\bibfnamefont {B.~K.}\ \bibnamefont
  {Agrawal}}, \bibinfo {author} {\bibfnamefont {G.}~\bibnamefont {Colo}},
  \bibinfo {author} {\bibfnamefont {N.}~\bibnamefont {Paar}}, \bibinfo {author}
  {\bibfnamefont {J.}~\bibnamefont {Piekarewicz}},\ and\ \bibinfo {author}
  {\bibfnamefont {D.}~\bibnamefont {Vretenar}},\ }\href@noop {} {\bibfield
  {journal} {\bibinfo  {journal} {Physical Review C}\ }\textbf {\bibinfo
  {volume} {92}},\ \bibinfo {pages} {064304} (\bibinfo {year}
  {2015})}\BibitemShut {NoStop}%
\bibitem [{\citenamefont {Hashimoto}\ \emph {et~al.}(2015)\citenamefont
  {Hashimoto}, \citenamefont {Krumbholz}, \citenamefont {Reinhard},
  \citenamefont {Tamii}, \citenamefont {von Neumann-Cosel}, \citenamefont
  {Adachi}, \citenamefont {Aoi}, \citenamefont {Bertulani}, \citenamefont
  {Fujita}, \citenamefont {Fujita} \emph {et~al.}}]{hashimoto2015dipole}%
  \BibitemOpen
  \bibfield  {author} {\bibinfo {author} {\bibfnamefont {T.}~\bibnamefont
  {Hashimoto}}, \bibinfo {author} {\bibfnamefont {A.~M.}\ \bibnamefont
  {Krumbholz}}, \bibinfo {author} {\bibfnamefont {P.~G.}\ \bibnamefont
  {Reinhard}}, \bibinfo {author} {\bibfnamefont {A.}~\bibnamefont {Tamii}},
  \bibinfo {author} {\bibfnamefont {P.}~\bibnamefont {von Neumann-Cosel}},
  \bibinfo {author} {\bibfnamefont {T.}~\bibnamefont {Adachi}}, \bibinfo
  {author} {\bibfnamefont {N.}~\bibnamefont {Aoi}}, \bibinfo {author}
  {\bibfnamefont {C.~A.}\ \bibnamefont {Bertulani}}, \bibinfo {author}
  {\bibfnamefont {H.}~\bibnamefont {Fujita}}, \bibinfo {author} {\bibfnamefont
  {Y.}~\bibnamefont {Fujita}}, \emph {et~al.},\ }\href@noop {} {\bibfield
  {journal} {\bibinfo  {journal} {Physical Review C}\ }\textbf {\bibinfo
  {volume} {92}},\ \bibinfo {pages} {031305} (\bibinfo {year}
  {2015})}\BibitemShut {NoStop}%
\bibitem [{\citenamefont {Roca-Maza}\ and\ \citenamefont
  {Paar}(2018)}]{roca2018nuclear}%
  \BibitemOpen
  \bibfield  {author} {\bibinfo {author} {\bibfnamefont {X.}~\bibnamefont
  {Roca-Maza}}\ and\ \bibinfo {author} {\bibfnamefont {N.}~\bibnamefont
  {Paar}},\ }\href@noop {} {\bibfield  {journal} {\bibinfo  {journal} {Progress
  in Particle and Nuclear Physics}\ }\textbf {\bibinfo {volume} {101}},\
  \bibinfo {pages} {96} (\bibinfo {year} {2018})}\BibitemShut {NoStop}%
\bibitem [{\citenamefont {Bassauer}\ \emph {et~al.}(2020)\citenamefont
  {Bassauer}, \citenamefont {von Neumann-Cosel}, \citenamefont {Reinhard},
  \citenamefont {Tamii}, \citenamefont {Adachi}, \citenamefont {Bertulani},
  \citenamefont {Chan}, \citenamefont {Col{\`o}}, \citenamefont {D'Alessio},
  \citenamefont {Fujioka} \emph {et~al.}}]{bassauer2020evolution}%
  \BibitemOpen
  \bibfield  {author} {\bibinfo {author} {\bibfnamefont {S.}~\bibnamefont
  {Bassauer}}, \bibinfo {author} {\bibfnamefont {P.}~\bibnamefont {von
  Neumann-Cosel}}, \bibinfo {author} {\bibfnamefont {P.-G.}\ \bibnamefont
  {Reinhard}}, \bibinfo {author} {\bibfnamefont {A.}~\bibnamefont {Tamii}},
  \bibinfo {author} {\bibfnamefont {S.}~\bibnamefont {Adachi}}, \bibinfo
  {author} {\bibfnamefont {C.~A.}\ \bibnamefont {Bertulani}}, \bibinfo {author}
  {\bibfnamefont {P.}~\bibnamefont {Chan}}, \bibinfo {author} {\bibfnamefont
  {G.}~\bibnamefont {Col{\`o}}}, \bibinfo {author} {\bibfnamefont
  {A.}~\bibnamefont {D'Alessio}}, \bibinfo {author} {\bibfnamefont
  {H.}~\bibnamefont {Fujioka}}, \emph {et~al.},\ }\href@noop {} {\bibfield
  {journal} {\bibinfo  {journal} {Physics Letters B}\ }\textbf {\bibinfo
  {volume} {810}},\ \bibinfo {pages} {135804} (\bibinfo {year}
  {2020})}\BibitemShut {NoStop}%
\bibitem [{\citenamefont {Orce}(2020)}]{orce2020polarizability}%
  \BibitemOpen
  \bibfield  {author} {\bibinfo {author} {\bibfnamefont {J.~N.}\ \bibnamefont
  {Orce}},\ }\href@noop {} {\bibfield  {journal} {\bibinfo  {journal}
  {International Journal of Modern Physics E}\ }\textbf {\bibinfo {volume}
  {29}},\ \bibinfo {pages} {2030002} (\bibinfo {year} {2020})}\BibitemShut
  {NoStop}%
\bibitem [{\citenamefont {Mott}\ and\ \citenamefont
  {Sneddon}(1948)}]{mott1948}%
  \BibitemOpen
  \bibfield  {author} {\bibinfo {author} {\bibfnamefont {N.~F.}\ \bibnamefont
  {Mott}}\ and\ \bibinfo {author} {\bibfnamefont {T.~N.}\ \bibnamefont
  {Sneddon}},\ }\href@noop {} {\emph {\bibinfo {title} {Wave Mechanics and its
  Applications}}}\ (\bibinfo  {publisher} {Clarendon Press, Oxford},\ \bibinfo
  {year} {1948})\BibitemShut {NoStop}%
\bibitem [{\citenamefont {Orce}\ \emph
  {et~al.}(2023{\natexlab{a}})\citenamefont {Orce}, \citenamefont {Dey},
  \citenamefont {Ngwetsheni}, \citenamefont {Bhattacharya}, \citenamefont
  {Pandit}, \citenamefont {Lesch},\ and\ \citenamefont
  {Zulu}}]{orce2021enhanced}%
  \BibitemOpen
  \bibfield  {author} {\bibinfo {author} {\bibfnamefont {J.~N.}\ \bibnamefont
  {Orce}}, \bibinfo {author} {\bibfnamefont {B.}~\bibnamefont {Dey}}, \bibinfo
  {author} {\bibfnamefont {C.}~\bibnamefont {Ngwetsheni}}, \bibinfo {author}
  {\bibfnamefont {S.}~\bibnamefont {Bhattacharya}}, \bibinfo {author}
  {\bibfnamefont {D.}~\bibnamefont {Pandit}}, \bibinfo {author} {\bibfnamefont
  {B.}~\bibnamefont {Lesch}},\ and\ \bibinfo {author} {\bibfnamefont
  {A.}~\bibnamefont {Zulu}},\ }\href@noop {} {\bibfield  {journal} {\bibinfo
  {journal} {Monthly Notices of the Royal Astronomical Society}\ }\textbf
  {\bibinfo {volume} {525}},\ \bibinfo {pages} {6249} (\bibinfo {year}
  {2023}{\natexlab{a}})}\BibitemShut {NoStop}%
\bibitem [{\citenamefont {Danos}(1958)}]{danos1958long}%
  \BibitemOpen
  \bibfield  {author} {\bibinfo {author} {\bibfnamefont {M.}~\bibnamefont
  {Danos}},\ }\href@noop {} {\bibfield  {journal} {\bibinfo  {journal} {Nuclear
  Physics}\ }\textbf {\bibinfo {volume} {5}},\ \bibinfo {pages} {23} (\bibinfo
  {year} {1958})}\BibitemShut {NoStop}%
\bibitem [{\citenamefont {Brink}(1955)}]{brink_1955}%
  \BibitemOpen
  \bibfield  {author} {\bibinfo {author} {\bibfnamefont {D.}~\bibnamefont
  {Brink}},\ }\href@noop {} {\emph {\bibinfo {title} {Some Aspects of the
  Interaction of Light with Matter}}}\ (\bibinfo  {publisher} {University of
  Oxford},\ \bibinfo {year} {1955})\BibitemShut {NoStop}%
\bibitem [{\citenamefont {Axel}(1962)}]{axel1962electric}%
  \BibitemOpen
  \bibfield  {author} {\bibinfo {author} {\bibfnamefont {P.}~\bibnamefont
  {Axel}},\ }\href@noop {} {\bibfield  {journal} {\bibinfo  {journal} {Physical
  Review}\ }\textbf {\bibinfo {volume} {126}},\ \bibinfo {pages} {671}
  (\bibinfo {year} {1962})}\BibitemShut {NoStop}%
\bibitem [{\citenamefont {Kusnezov}\ \emph {et~al.}(1998)\citenamefont
  {Kusnezov}, \citenamefont {Alhassid},\ and\ \citenamefont
  {Snover}}]{kusnezov1998scaling}%
  \BibitemOpen
  \bibfield  {author} {\bibinfo {author} {\bibfnamefont {D.}~\bibnamefont
  {Kusnezov}}, \bibinfo {author} {\bibfnamefont {Y.}~\bibnamefont {Alhassid}},\
  and\ \bibinfo {author} {\bibfnamefont {K.~A.}\ \bibnamefont {Snover}},\
  }\href@noop {} {\bibfield  {journal} {\bibinfo  {journal} {Physical review
  letters}\ }\textbf {\bibinfo {volume} {81}},\ \bibinfo {pages} {542}
  (\bibinfo {year} {1998})}\BibitemShut {NoStop}%
\bibitem [{\citenamefont {Snover}(1986)}]{snover1986giant}%
  \BibitemOpen
  \bibfield  {author} {\bibinfo {author} {\bibfnamefont {K.~A.}\ \bibnamefont
  {Snover}},\ }\href@noop {} {\bibfield  {journal} {\bibinfo  {journal} {Annual
  Review of Nuclear and Particle Science}\ }\textbf {\bibinfo {volume} {36}},\
  \bibinfo {pages} {545} (\bibinfo {year} {1986})}\BibitemShut {NoStop}%
\bibitem [{\citenamefont {Gaardhoje}(1992)}]{gaardhoje1992nuclear}%
  \BibitemOpen
  \bibfield  {author} {\bibinfo {author} {\bibfnamefont {J.}~\bibnamefont
  {Gaardhoje}},\ }\href@noop {} {\bibfield  {journal} {\bibinfo  {journal}
  {Annual Review of Nuclear and Particle Science}\ }\textbf {\bibinfo {volume}
  {42}},\ \bibinfo {pages} {483} (\bibinfo {year} {1992})}\BibitemShut
  {NoStop}%
\bibitem [{\citenamefont {Schiller}\ and\ \citenamefont
  {Thoennessen}(2007)}]{schiller2007compilation}%
  \BibitemOpen
  \bibfield  {author} {\bibinfo {author} {\bibfnamefont {A.}~\bibnamefont
  {Schiller}}\ and\ \bibinfo {author} {\bibfnamefont {M.}~\bibnamefont
  {Thoennessen}},\ }\href@noop {} {\bibfield  {journal} {\bibinfo  {journal}
  {Atomic Data and Nuclear Data Tables}\ }\textbf {\bibinfo {volume} {93}},\
  \bibinfo {pages} {549} (\bibinfo {year} {2007})}\BibitemShut {NoStop}%
\bibitem [{\citenamefont {Bracco}\ and\ \citenamefont
  {Camera}(1998)}]{bracco1998status}%
  \BibitemOpen
  \bibfield  {author} {\bibinfo {author} {\bibfnamefont {A.}~\bibnamefont
  {Bracco}}\ and\ \bibinfo {author} {\bibfnamefont {F.}~\bibnamefont
  {Camera}},\ }\href@noop {} {\bibfield  {journal} {\bibinfo  {journal} {Il
  Nuovo Cimento A (1971-1996)}\ }\textbf {\bibinfo {volume} {111}},\ \bibinfo
  {pages} {607} (\bibinfo {year} {1998})}\BibitemShut {NoStop}%
\bibitem [{\citenamefont {Mondal}\ \emph {et~al.}(2018)\citenamefont {Mondal},
  \citenamefont {Pandit}, \citenamefont {Mukhopadhyay}, \citenamefont {Pal},
  \citenamefont {Bhattacharya}, \citenamefont {De}, \citenamefont {Dang},
  \citenamefont {Hung}, \citenamefont {Bhattacharya}, \citenamefont
  {Bhattacharyya} \emph {et~al.}}]{mondal2018study}%
  \BibitemOpen
  \bibfield  {author} {\bibinfo {author} {\bibfnamefont {D.}~\bibnamefont
  {Mondal}}, \bibinfo {author} {\bibfnamefont {D.}~\bibnamefont {Pandit}},
  \bibinfo {author} {\bibfnamefont {S.}~\bibnamefont {Mukhopadhyay}}, \bibinfo
  {author} {\bibfnamefont {S.}~\bibnamefont {Pal}}, \bibinfo {author}
  {\bibfnamefont {S.}~\bibnamefont {Bhattacharya}}, \bibinfo {author}
  {\bibfnamefont {A.}~\bibnamefont {De}}, \bibinfo {author} {\bibfnamefont
  {N.~D.}\ \bibnamefont {Dang}}, \bibinfo {author} {\bibfnamefont {N.~Q.}\
  \bibnamefont {Hung}}, \bibinfo {author} {\bibfnamefont {S.}~\bibnamefont
  {Bhattacharya}}, \bibinfo {author} {\bibfnamefont {S.}~\bibnamefont
  {Bhattacharyya}}, \emph {et~al.},\ }\href@noop {} {\bibfield  {journal}
  {\bibinfo  {journal} {Physics Letters B}\ }\textbf {\bibinfo {volume}
  {784}},\ \bibinfo {pages} {423} (\bibinfo {year} {2018})}\BibitemShut
  {NoStop}%
\bibitem [{\citenamefont {Kici{\'n}ska-Habior}\ \emph
  {et~al.}(1987)\citenamefont {Kici{\'n}ska-Habior}, \citenamefont {Snover},
  \citenamefont {Gossett}, \citenamefont {Behr}, \citenamefont {Feldman},
  \citenamefont {Glatzel}, \citenamefont {Gundlach},\ and\ \citenamefont
  {Garman}}]{kicinska1987statistical}%
  \BibitemOpen
  \bibfield  {author} {\bibinfo {author} {\bibfnamefont {M.}~\bibnamefont
  {Kici{\'n}ska-Habior}}, \bibinfo {author} {\bibfnamefont {K.}~\bibnamefont
  {Snover}}, \bibinfo {author} {\bibfnamefont {C.}~\bibnamefont {Gossett}},
  \bibinfo {author} {\bibfnamefont {J.}~\bibnamefont {Behr}}, \bibinfo {author}
  {\bibfnamefont {G.}~\bibnamefont {Feldman}}, \bibinfo {author} {\bibfnamefont
  {H.}~\bibnamefont {Glatzel}}, \bibinfo {author} {\bibfnamefont
  {J.}~\bibnamefont {Gundlach}},\ and\ \bibinfo {author} {\bibfnamefont
  {E.}~\bibnamefont {Garman}},\ }\href@noop {} {\bibfield  {journal} {\bibinfo
  {journal} {Physical Review C}\ }\textbf {\bibinfo {volume} {36}},\ \bibinfo
  {pages} {612} (\bibinfo {year} {1987})}\BibitemShut {NoStop}%
\bibitem [{\citenamefont {Dey}\ \emph {et~al.}(2014)\citenamefont {Dey},
  \citenamefont {Mondal}, \citenamefont {Pandit}, \citenamefont {Mukhopadhyay},
  \citenamefont {Pal}, \citenamefont {Bhattacharya}, \citenamefont {De},
  \citenamefont {Banerjee}, \citenamefont {Dang}, \citenamefont {Hung} \emph
  {et~al.}}]{dey2014probing}%
  \BibitemOpen
  \bibfield  {author} {\bibinfo {author} {\bibfnamefont {B.}~\bibnamefont
  {Dey}}, \bibinfo {author} {\bibfnamefont {D.}~\bibnamefont {Mondal}},
  \bibinfo {author} {\bibfnamefont {D.}~\bibnamefont {Pandit}}, \bibinfo
  {author} {\bibfnamefont {S.}~\bibnamefont {Mukhopadhyay}}, \bibinfo {author}
  {\bibfnamefont {S.}~\bibnamefont {Pal}}, \bibinfo {author} {\bibfnamefont
  {S.}~\bibnamefont {Bhattacharya}}, \bibinfo {author} {\bibfnamefont
  {A.}~\bibnamefont {De}}, \bibinfo {author} {\bibfnamefont {K.}~\bibnamefont
  {Banerjee}}, \bibinfo {author} {\bibfnamefont {N.~D.}\ \bibnamefont {Dang}},
  \bibinfo {author} {\bibfnamefont {N.~Q.}\ \bibnamefont {Hung}}, \emph
  {et~al.},\ }\href@noop {} {\bibfield  {journal} {\bibinfo  {journal} {Physics
  Letters B}\ }\textbf {\bibinfo {volume} {731}},\ \bibinfo {pages} {92}
  (\bibinfo {year} {2014})}\BibitemShut {NoStop}%
\bibitem [{\citenamefont {Heckman}\ \emph {et~al.}(2003)\citenamefont
  {Heckman}, \citenamefont {Bazin}, \citenamefont {Beene}, \citenamefont
  {Blumenfeld}, \citenamefont {Chromik}, \citenamefont {Halbert}, \citenamefont
  {Liang}, \citenamefont {Mohrmann}, \citenamefont {Nakamura}, \citenamefont
  {Navin} \emph {et~al.}}]{heckman2003low}%
  \BibitemOpen
  \bibfield  {author} {\bibinfo {author} {\bibfnamefont {P.}~\bibnamefont
  {Heckman}}, \bibinfo {author} {\bibfnamefont {D.}~\bibnamefont {Bazin}},
  \bibinfo {author} {\bibfnamefont {J.}~\bibnamefont {Beene}}, \bibinfo
  {author} {\bibfnamefont {Y.}~\bibnamefont {Blumenfeld}}, \bibinfo {author}
  {\bibfnamefont {M.}~\bibnamefont {Chromik}}, \bibinfo {author} {\bibfnamefont
  {M.}~\bibnamefont {Halbert}}, \bibinfo {author} {\bibfnamefont
  {J.}~\bibnamefont {Liang}}, \bibinfo {author} {\bibfnamefont
  {E.}~\bibnamefont {Mohrmann}}, \bibinfo {author} {\bibfnamefont
  {T.}~\bibnamefont {Nakamura}}, \bibinfo {author} {\bibfnamefont
  {A.}~\bibnamefont {Navin}}, \emph {et~al.},\ }\href@noop {} {\bibfield
  {journal} {\bibinfo  {journal} {Physics Letters B}\ }\textbf {\bibinfo
  {volume} {555}},\ \bibinfo {pages} {43} (\bibinfo {year} {2003})}\BibitemShut
  {NoStop}%
\bibitem [{\citenamefont {Pandit}\ \emph {et~al.}(2012)\citenamefont {Pandit},
  \citenamefont {Mukhopadhyay}, \citenamefont {Pal}, \citenamefont {De},\ and\
  \citenamefont {Banerjee}}]{pandit2012critical}%
  \BibitemOpen
  \bibfield  {author} {\bibinfo {author} {\bibfnamefont {D.}~\bibnamefont
  {Pandit}}, \bibinfo {author} {\bibfnamefont {S.}~\bibnamefont
  {Mukhopadhyay}}, \bibinfo {author} {\bibfnamefont {S.}~\bibnamefont {Pal}},
  \bibinfo {author} {\bibfnamefont {A.}~\bibnamefont {De}},\ and\ \bibinfo
  {author} {\bibfnamefont {S.}~\bibnamefont {Banerjee}},\ }\href@noop {}
  {\bibfield  {journal} {\bibinfo  {journal} {Physics Letters B}\ }\textbf
  {\bibinfo {volume} {713}},\ \bibinfo {pages} {434} (\bibinfo {year}
  {2012})}\BibitemShut {NoStop}%
\bibitem [{\citenamefont {Gossett}\ \emph {et~al.}(1985)\citenamefont
  {Gossett}, \citenamefont {Snover}, \citenamefont {Behr}, \citenamefont
  {Feldman},\ and\ \citenamefont {Osborne}}]{gossett1985deformation}%
  \BibitemOpen
  \bibfield  {author} {\bibinfo {author} {\bibfnamefont {C.}~\bibnamefont
  {Gossett}}, \bibinfo {author} {\bibfnamefont {K.}~\bibnamefont {Snover}},
  \bibinfo {author} {\bibfnamefont {J.}~\bibnamefont {Behr}}, \bibinfo {author}
  {\bibfnamefont {G.}~\bibnamefont {Feldman}},\ and\ \bibinfo {author}
  {\bibfnamefont {J.}~\bibnamefont {Osborne}},\ }\href@noop {} {\bibfield
  {journal} {\bibinfo  {journal} {Physical Review Letters}\ }\textbf {\bibinfo
  {volume} {54}},\ \bibinfo {pages} {1486} (\bibinfo {year}
  {1985})}\BibitemShut {NoStop}%
\bibitem [{\citenamefont {Pandit}\ \emph {et~al.}(2021)\citenamefont {Pandit},
  \citenamefont {Dey}, \citenamefont {Bhattacharya}, \citenamefont {Rana},
  \citenamefont {Mondal}, \citenamefont {Mukhopadhyay}, \citenamefont {Pal},
  \citenamefont {De}, \citenamefont {Roy}, \citenamefont {Banerjee} \emph
  {et~al.}}]{pandit2021puzzle}%
  \BibitemOpen
  \bibfield  {author} {\bibinfo {author} {\bibfnamefont {D.}~\bibnamefont
  {Pandit}}, \bibinfo {author} {\bibfnamefont {B.}~\bibnamefont {Dey}},
  \bibinfo {author} {\bibfnamefont {S.}~\bibnamefont {Bhattacharya}}, \bibinfo
  {author} {\bibfnamefont {T.}~\bibnamefont {Rana}}, \bibinfo {author}
  {\bibfnamefont {D.}~\bibnamefont {Mondal}}, \bibinfo {author} {\bibfnamefont
  {S.}~\bibnamefont {Mukhopadhyay}}, \bibinfo {author} {\bibfnamefont
  {S.}~\bibnamefont {Pal}}, \bibinfo {author} {\bibfnamefont {A.}~\bibnamefont
  {De}}, \bibinfo {author} {\bibfnamefont {P.}~\bibnamefont {Roy}}, \bibinfo
  {author} {\bibfnamefont {K.}~\bibnamefont {Banerjee}}, \emph {et~al.},\
  }\href@noop {} {\bibfield  {journal} {\bibinfo  {journal} {Physics Letters
  B}\ }\textbf {\bibinfo {volume} {816}},\ \bibinfo {pages} {136173} (\bibinfo
  {year} {2021})}\BibitemShut {NoStop}%
\bibitem [{\citenamefont {H{\"a}usser}\ \emph {et~al.}(1973)\citenamefont
  {H{\"a}usser}, \citenamefont {McDonald}, \citenamefont {Alexander},
  \citenamefont {Ferguson},\ and\ \citenamefont {Warner}}]{hausser1973e1}%
  \BibitemOpen
  \bibfield  {author} {\bibinfo {author} {\bibfnamefont {O.}~\bibnamefont
  {H{\"a}usser}}, \bibinfo {author} {\bibfnamefont {A.~B.}\ \bibnamefont
  {McDonald}}, \bibinfo {author} {\bibfnamefont {T.~K.}\ \bibnamefont
  {Alexander}}, \bibinfo {author} {\bibfnamefont {A.~J.}\ \bibnamefont
  {Ferguson}},\ and\ \bibinfo {author} {\bibfnamefont {R.~E.}\ \bibnamefont
  {Warner}},\ }\href@noop {} {\bibfield  {journal} {\bibinfo  {journal}
  {Nuclear Physics A}\ }\textbf {\bibinfo {volume} {212}},\ \bibinfo {pages}
  {613} (\bibinfo {year} {1973})}\BibitemShut {NoStop}%
\bibitem [{\citenamefont {Vermeer}\ \emph
  {et~al.}(1984{\natexlab{a}})\citenamefont {Vermeer}, \citenamefont {Baxter},
  \citenamefont {Burnett}, \citenamefont {Esat}, \citenamefont {Fewell},\ and\
  \citenamefont {Spear}}]{vermeer1984coulomb}%
  \BibitemOpen
  \bibfield  {author} {\bibinfo {author} {\bibfnamefont {W.~J.}\ \bibnamefont
  {Vermeer}}, \bibinfo {author} {\bibfnamefont {A.~M.}\ \bibnamefont {Baxter}},
  \bibinfo {author} {\bibfnamefont {S.~M.}\ \bibnamefont {Burnett}}, \bibinfo
  {author} {\bibfnamefont {M.~T.}\ \bibnamefont {Esat}}, \bibinfo {author}
  {\bibfnamefont {M.~P.}\ \bibnamefont {Fewell}},\ and\ \bibinfo {author}
  {\bibfnamefont {R.~H.}\ \bibnamefont {Spear}},\ }\href@noop {} {\bibfield
  {journal} {\bibinfo  {journal} {Australian Journal of Physics}\ }\textbf
  {\bibinfo {volume} {37}},\ \bibinfo {pages} {273} (\bibinfo {year}
  {1984}{\natexlab{a}})}\BibitemShut {NoStop}%
\bibitem [{\citenamefont {Vermeer}\ \emph
  {et~al.}(1984{\natexlab{b}})\citenamefont {Vermeer}, \citenamefont {Esat},
  \citenamefont {Fewell}, \citenamefont {Spear}, \citenamefont {Baxter},\ and\
  \citenamefont {Burnett}}]{vermeer1984experimental}%
  \BibitemOpen
  \bibfield  {author} {\bibinfo {author} {\bibfnamefont {W.~J.}\ \bibnamefont
  {Vermeer}}, \bibinfo {author} {\bibfnamefont {M.~T.}\ \bibnamefont {Esat}},
  \bibinfo {author} {\bibfnamefont {M.~P.}\ \bibnamefont {Fewell}}, \bibinfo
  {author} {\bibfnamefont {R.~H.}\ \bibnamefont {Spear}}, \bibinfo {author}
  {\bibfnamefont {A.~M.}\ \bibnamefont {Baxter}},\ and\ \bibinfo {author}
  {\bibfnamefont {S.~M.}\ \bibnamefont {Burnett}},\ }\href@noop {} {\bibfield
  {journal} {\bibinfo  {journal} {Physics Letters B}\ }\textbf {\bibinfo
  {volume} {138}},\ \bibinfo {pages} {365} (\bibinfo {year}
  {1984}{\natexlab{b}})}\BibitemShut {NoStop}%
\bibitem [{\citenamefont {Kuehner}\ \emph {et~al.}(1982)\citenamefont
  {Kuehner}, \citenamefont {Spear}, \citenamefont {Vermeer}, \citenamefont
  {Esat}, \citenamefont {Baxter},\ and\ \citenamefont
  {Hinds}}]{kuehner1982measurement}%
  \BibitemOpen
  \bibfield  {author} {\bibinfo {author} {\bibfnamefont {J.~A.}\ \bibnamefont
  {Kuehner}}, \bibinfo {author} {\bibfnamefont {R.~H.}\ \bibnamefont {Spear}},
  \bibinfo {author} {\bibfnamefont {W.~J.}\ \bibnamefont {Vermeer}}, \bibinfo
  {author} {\bibfnamefont {M.~T.}\ \bibnamefont {Esat}}, \bibinfo {author}
  {\bibfnamefont {A.~M.}\ \bibnamefont {Baxter}},\ and\ \bibinfo {author}
  {\bibfnamefont {S.}~\bibnamefont {Hinds}},\ }\href@noop {} {\bibfield
  {journal} {\bibinfo  {journal} {Physics Letters B}\ }\textbf {\bibinfo
  {volume} {115}},\ \bibinfo {pages} {437} (\bibinfo {year}
  {1982})}\BibitemShut {NoStop}%
\bibitem [{\citenamefont {Alder}\ and\ \citenamefont
  {Winther}(1975)}]{alder1975north}%
  \BibitemOpen
  \bibfield  {author} {\bibinfo {author} {\bibfnamefont {K.}~\bibnamefont
  {Alder}}\ and\ \bibinfo {author} {\bibfnamefont {A.}~\bibnamefont
  {Winther}},\ }\href@noop {} {\emph {\bibinfo {title} {Electromagnetic
  Excitation}}}\ (\bibinfo  {publisher} {North-Holland},\ \bibinfo {address}
  {Amsterdam},\ \bibinfo {year} {1975})\BibitemShut {NoStop}%
\bibitem [{\citenamefont {DeShalit}\ and\ \citenamefont
  {Feshbach}(1974)}]{deshalit1974theoretical}%
  \BibitemOpen
  \bibfield  {author} {\bibinfo {author} {\bibfnamefont {A.}~\bibnamefont
  {DeShalit}}\ and\ \bibinfo {author} {\bibfnamefont {H.}~\bibnamefont
  {Feshbach}},\ }\href@noop {} {\emph {\bibinfo {title} {Theoretical nuclear
  physics. Volume I. Nuclear structure}}}\ (\bibinfo  {publisher} {John Wiley
  and Sons, Inc., New York},\ \bibinfo {year} {1974})\BibitemShut {NoStop}%
\bibitem [{\citenamefont {Suhonen}(2007)}]{suhonen2007nucleons}%
  \BibitemOpen
  \bibfield  {author} {\bibinfo {author} {\bibfnamefont {J.}~\bibnamefont
  {Suhonen}},\ }\href@noop {} {\emph {\bibinfo {title} {From nucleons to
  nucleus: concepts of microscopic nuclear theory}}}\ (\bibinfo  {publisher}
  {Springer Science \& Business Media},\ \bibinfo {year} {2007})\BibitemShut
  {NoStop}%
\bibitem [{\citenamefont {Eichler}(1964)}]{eichler1964second}%
  \BibitemOpen
  \bibfield  {author} {\bibinfo {author} {\bibfnamefont {J.}~\bibnamefont
  {Eichler}},\ }\href@noop {} {\bibfield  {journal} {\bibinfo  {journal}
  {Physical Review}\ }\textbf {\bibinfo {volume} {133}} (\bibinfo {year}
  {1964})}\BibitemShut {NoStop}%
\bibitem [{\citenamefont {de~Boer}\ and\ \citenamefont
  {Eichler}(1968)}]{deBoer1968}%
  \BibitemOpen
  \bibfield  {author} {\bibinfo {author} {\bibfnamefont {J.}~\bibnamefont
  {de~Boer}}\ and\ \bibinfo {author} {\bibfnamefont {J.}~\bibnamefont
  {Eichler}},\ }\href@noop {} {\bibfield  {journal} {\bibinfo  {journal}
  {Advances in Nuclear Physics}\ }\textbf {\bibinfo {volume} {1}},\ \bibinfo
  {pages} {1} (\bibinfo {year} {1968})}\BibitemShut {NoStop}%
\bibitem [{\citenamefont {H{\"a}usser}(1974)}]{hausser1974}%
  \BibitemOpen
  \bibfield  {author} {\bibinfo {author} {\bibfnamefont {O.}~\bibnamefont
  {H{\"a}usser}},\ }\href@noop {} {\bibfield  {journal} {\bibinfo  {journal}
  {Nuclear Spectroscopy and Reactions}\ }\textbf {\bibinfo {volume} {C}},\
  \bibinfo {pages} {55} (\bibinfo {year} {1974})}\BibitemShut {NoStop}%
\bibitem [{\citenamefont {Flambaum}\ \emph {et~al.}(2021)\citenamefont
  {Flambaum}, \citenamefont {Samsonov}, \citenamefont {Tan},\ and\
  \citenamefont {Viatkina}}]{flambaum2021nuclear}%
  \BibitemOpen
  \bibfield  {author} {\bibinfo {author} {\bibfnamefont {V.~V.}\ \bibnamefont
  {Flambaum}}, \bibinfo {author} {\bibfnamefont {I.~B.}\ \bibnamefont
  {Samsonov}}, \bibinfo {author} {\bibfnamefont {H.~B.~T.}\ \bibnamefont
  {Tan}},\ and\ \bibinfo {author} {\bibfnamefont {A.~V.}\ \bibnamefont
  {Viatkina}},\ }\href@noop {} {\bibfield  {journal} {\bibinfo  {journal}
  {Physical Review A}\ }\textbf {\bibinfo {volume} {103}},\ \bibinfo {pages}
  {032811} (\bibinfo {year} {2021})}\BibitemShut {NoStop}%
\bibitem [{\citenamefont {Barker}(1982{\natexlab{a}})}]{barger1982gdr}%
  \BibitemOpen
  \bibfield  {author} {\bibinfo {author} {\bibfnamefont {F.~C.}\ \bibnamefont
  {Barker}},\ }\href@noop {} {\bibfield  {journal} {\bibinfo  {journal}
  {Australian Journal of Physics}\ }\textbf {\bibinfo {volume} {35}},\ \bibinfo
  {pages} {291} (\bibinfo {year} {1982}{\natexlab{a}})}\BibitemShut {NoStop}%
\bibitem [{\citenamefont {Barker}(1982{\natexlab{b}})}]{barker1982gdr17O}%
  \BibitemOpen
  \bibfield  {author} {\bibinfo {author} {\bibfnamefont {F.~C.}\ \bibnamefont
  {Barker}},\ }\href@noop {} {\bibfield  {journal} {\bibinfo  {journal}
  {Australian Journal of Physics}\ }\textbf {\bibinfo {volume} {35}},\ \bibinfo
  {pages} {301} (\bibinfo {year} {1982}{\natexlab{b}})}\BibitemShut {NoStop}%
\bibitem [{\citenamefont {Barker}(1982{\natexlab{c}})}]{barker1982gdr18O}%
  \BibitemOpen
  \bibfield  {author} {\bibinfo {author} {\bibfnamefont {F.~C.}\ \bibnamefont
  {Barker}},\ }\href@noop {} {\bibfield  {journal} {\bibinfo  {journal}
  {Australian Journal of Physics}\ }\textbf {\bibinfo {volume} {35}},\ \bibinfo
  {pages} {377} (\bibinfo {year} {1982}{\natexlab{c}})}\BibitemShut {NoStop}%
\bibitem [{\citenamefont {Barker}\ and\ \citenamefont
  {Woods}(1989)}]{barke1989investigation}%
  \BibitemOpen
  \bibfield  {author} {\bibinfo {author} {\bibfnamefont {F.~C.}\ \bibnamefont
  {Barker}}\ and\ \bibinfo {author} {\bibfnamefont {C.~L.}\ \bibnamefont
  {Woods}},\ }\href@noop {} {\bibfield  {journal} {\bibinfo  {journal}
  {Australian Journal of Physics}\ }\textbf {\bibinfo {volume} {42}},\ \bibinfo
  {pages} {233} (\bibinfo {year} {1989})}\BibitemShut {NoStop}%
\bibitem [{\citenamefont {Kumar-Raju}\ \emph {et~al.}(2018)\citenamefont
  {Kumar-Raju}, \citenamefont {Orce}, \citenamefont {Navr{\'a}til},
  \citenamefont {Ball}, \citenamefont {Drake}, \citenamefont {Triambak},
  \citenamefont {Hackman}, \citenamefont {Pearson}, \citenamefont {Abrahams},
  \citenamefont {Akakpo} \emph {et~al.}}]{raju2018reorientation}%
  \BibitemOpen
  \bibfield  {author} {\bibinfo {author} {\bibfnamefont {M.}~\bibnamefont
  {Kumar-Raju}}, \bibinfo {author} {\bibfnamefont {J.~N.}\ \bibnamefont
  {Orce}}, \bibinfo {author} {\bibfnamefont {P.}~\bibnamefont {Navr{\'a}til}},
  \bibinfo {author} {\bibfnamefont {G.~C.}\ \bibnamefont {Ball}}, \bibinfo
  {author} {\bibfnamefont {T.~E.}\ \bibnamefont {Drake}}, \bibinfo {author}
  {\bibfnamefont {S.}~\bibnamefont {Triambak}}, \bibinfo {author}
  {\bibfnamefont {G.}~\bibnamefont {Hackman}}, \bibinfo {author} {\bibfnamefont
  {C.~J.}\ \bibnamefont {Pearson}}, \bibinfo {author} {\bibfnamefont {K.~J.}\
  \bibnamefont {Abrahams}}, \bibinfo {author} {\bibfnamefont {E.~H.}\
  \bibnamefont {Akakpo}}, \emph {et~al.},\ }\href@noop {} {\bibfield  {journal}
  {\bibinfo  {journal} {Physics Letters B}\ }\textbf {\bibinfo {volume}
  {777}},\ \bibinfo {pages} {250} (\bibinfo {year} {2018})}\BibitemShut
  {NoStop}%
\bibitem [{\citenamefont {Orce}\ \emph
  {et~al.}(2023{\natexlab{b}})\citenamefont {Orce}, \citenamefont
  {Ngwetsheni},\ and\ \citenamefont {Brown}}]{orce2023global}%
  \BibitemOpen
  \bibfield  {author} {\bibinfo {author} {\bibfnamefont {J.~N.}\ \bibnamefont
  {Orce}}, \bibinfo {author} {\bibfnamefont {C.}~\bibnamefont {Ngwetsheni}},\
  and\ \bibinfo {author} {\bibfnamefont {B.~A.}\ \bibnamefont {Brown}},\
  }\href@noop {} {\bibfield  {journal} {\bibinfo  {journal} {Physical Review
  C}\ }\textbf {\bibinfo {volume} {108}},\ \bibinfo {pages} {044309} (\bibinfo
  {year} {2023}{\natexlab{b}})}\BibitemShut {NoStop}%
\bibitem [{\citenamefont {Warburton}\ and\ \citenamefont
  {Weneser}(1969)}]{warburton1969role}%
  \BibitemOpen
  \bibfield  {author} {\bibinfo {author} {\bibfnamefont {E.~K.}\ \bibnamefont
  {Warburton}}\ and\ \bibinfo {author} {\bibfnamefont {J.}~\bibnamefont
  {Weneser}},\ }\href@noop {} {\bibfield  {journal} {\bibinfo  {journal}
  {Isospin in Nuclear Physics, ed D. H. Wilkinson, North-Holland, Amsterdam}\
  }\textbf {\bibinfo {volume} {4}},\ \bibinfo {pages} {10} (\bibinfo {year}
  {1969})}\BibitemShut {NoStop}%
\bibitem [{\citenamefont {Rose}(1957)}]{rose1957elementary}%
  \BibitemOpen
  \bibfield  {author} {\bibinfo {author} {\bibfnamefont {M.~E.}\ \bibnamefont
  {Rose}},\ }\href@noop {} {\bibfield  {journal} {\bibinfo  {journal} {New
  York}\ } (\bibinfo {year} {1957})}\BibitemShut {NoStop}%
\bibitem [{\citenamefont {Messiah}(1961)}]{messiah1961quantum}%
  \BibitemOpen
  \bibfield  {author} {\bibinfo {author} {\bibfnamefont {A.}~\bibnamefont
  {Messiah}},\ }\href@noop {} {\bibinfo {title} {Quantum mechanics vol 1 \& 2
  tr. gm temmer}} (\bibinfo {year} {1961})\BibitemShut {NoStop}%
\bibitem [{\citenamefont {Levinger}(1960)}]{levinger1960}%
  \BibitemOpen
  \bibfield  {author} {\bibinfo {author} {\bibfnamefont {J.~S.}\ \bibnamefont
  {Levinger}},\ }\href@noop {} {\emph {\bibinfo {title} {Nuclear
  Photo-Disintegration}}}\ (\bibinfo  {publisher} {Oxford University Press},\
  \bibinfo {address} {Oxford},\ \bibinfo {year} {1960})\BibitemShut {NoStop}%
\bibitem [{\citenamefont {Migdal}\ \emph {et~al.}(1965)\citenamefont {Migdal},
  \citenamefont {Lushnikov},\ and\ \citenamefont
  {Zaretsky}}]{migdal1965theory}%
  \BibitemOpen
  \bibfield  {author} {\bibinfo {author} {\bibfnamefont {A.}~\bibnamefont
  {Migdal}}, \bibinfo {author} {\bibfnamefont {A.}~\bibnamefont {Lushnikov}},\
  and\ \bibinfo {author} {\bibfnamefont {D.}~\bibnamefont {Zaretsky}},\
  }\href@noop {} {\bibfield  {journal} {\bibinfo  {journal} {Nuclear Physics}\
  }\textbf {\bibinfo {volume} {66}},\ \bibinfo {pages} {193} (\bibinfo {year}
  {1965})}\BibitemShut {NoStop}%
\bibitem [{\citenamefont {Ahrens}\ \emph {et~al.}(1976)\citenamefont {Ahrens},
  \citenamefont {Gimm}, \citenamefont {Zieger},\ and\ \citenamefont
  {Ziegler}}]{ahrens1976experimental}%
  \BibitemOpen
  \bibfield  {author} {\bibinfo {author} {\bibfnamefont {J.}~\bibnamefont
  {Ahrens}}, \bibinfo {author} {\bibfnamefont {H.}~\bibnamefont {Gimm}},
  \bibinfo {author} {\bibfnamefont {A.}~\bibnamefont {Zieger}},\ and\ \bibinfo
  {author} {\bibfnamefont {B.}~\bibnamefont {Ziegler}},\ }\href@noop {}
  {\bibfield  {journal} {\bibinfo  {journal} {Il Nuovo Cimento A (1965-1970)}\
  }\textbf {\bibinfo {volume} {32}},\ \bibinfo {pages} {364} (\bibinfo {year}
  {1976})}\BibitemShut {NoStop}%
\bibitem [{\citenamefont {Orce}(2015)}]{orce2015new}%
  \BibitemOpen
  \bibfield  {author} {\bibinfo {author} {\bibfnamefont {J.~N.}\ \bibnamefont
  {Orce}},\ }\href@noop {} {\bibfield  {journal} {\bibinfo  {journal} {Physical
  Review C}\ }\textbf {\bibinfo {volume} {91}},\ \bibinfo {pages} {064602}
  (\bibinfo {year} {2015})}\BibitemShut {NoStop}%
\bibitem [{\citenamefont {Orce}(2016)}]{orce2016reply}%
  \BibitemOpen
  \bibfield  {author} {\bibinfo {author} {\bibfnamefont {J.~N.}\ \bibnamefont
  {Orce}},\ }\href@noop {} {\bibfield  {journal} {\bibinfo  {journal} {Physical
  Review C}\ }\textbf {\bibinfo {volume} {93}},\ \bibinfo {pages} {049802}
  (\bibinfo {year} {2016})}\BibitemShut {NoStop}%
\bibitem [{exf()}]{exfor}%
  \BibitemOpen
  \href@noop {} {\bibinfo {title} {{EXFOR}: Experimental nuclear reaction
  data}},\ \bibinfo {howpublished}
  {\url{https://www-nds.iaea.org/exfor/exfor.htm}},\ \bibinfo {note} {accessed:
  2022-05-21}\BibitemShut {NoStop}%
\bibitem [{END()}]{ENDF}%
  \BibitemOpen
  \href@noop {} {\bibinfo {title} {{ENDF}: Evaluated nuclear data file}},\
  \bibinfo {howpublished} {\url{https://www-nds.iaea.org/exfor/endf.htm}},\
  \bibinfo {note} {accessed: 2022-05-21}\BibitemShut {NoStop}%
\bibitem [{\citenamefont {Ngwetsheni}\ and\ \citenamefont
  {Orce}(2019{\natexlab{a}})}]{ngwetsheni2019continuing}%
  \BibitemOpen
  \bibfield  {author} {\bibinfo {author} {\bibfnamefont {C.}~\bibnamefont
  {Ngwetsheni}}\ and\ \bibinfo {author} {\bibfnamefont {J.~N.}\ \bibnamefont
  {Orce}},\ }\href@noop {} {\bibfield  {journal} {\bibinfo  {journal} {Physics
  Letters B}\ }\textbf {\bibinfo {volume} {792}},\ \bibinfo {pages} {335}
  (\bibinfo {year} {2019}{\natexlab{a}})}\BibitemShut {NoStop}%
\bibitem [{\citenamefont {Ngwetsheni}\ and\ \citenamefont
  {Orce}(2019{\natexlab{b}})}]{ngwetsheni2019combined}%
  \BibitemOpen
  \bibfield  {author} {\bibinfo {author} {\bibfnamefont {C.}~\bibnamefont
  {Ngwetsheni}}\ and\ \bibinfo {author} {\bibfnamefont {J.~N.}\ \bibnamefont
  {Orce}},\ }\href@noop {} {\bibfield  {journal} {\bibinfo  {journal}
  {Hyperfine Interactions}\ }\textbf {\bibinfo {volume} {240}},\ \bibinfo
  {pages} {94} (\bibinfo {year} {2019}{\natexlab{b}})}\BibitemShut {NoStop}%
\bibitem [{\citenamefont {Ngwetsheni}\ and\ \citenamefont
  {Orce}(2019{\natexlab{c}})}]{ngwetsheni2019how}%
  \BibitemOpen
  \bibfield  {author} {\bibinfo {author} {\bibfnamefont {C.}~\bibnamefont
  {Ngwetsheni}}\ and\ \bibinfo {author} {\bibfnamefont {J.~N.}\ \bibnamefont
  {Orce}},\ }\href@noop {} {\bibfield  {journal} {\bibinfo  {journal} {EPJ Web
  Conf.}\ }\textbf {\bibinfo {volume} {223}},\ \bibinfo {pages} {01045}
  (\bibinfo {year} {2019}{\natexlab{c}})}\BibitemShut {NoStop}%
\bibitem [{\citenamefont {Racah}(1942)}]{racah1942theory}%
  \BibitemOpen
  \bibfield  {author} {\bibinfo {author} {\bibfnamefont {G.}~\bibnamefont
  {Racah}},\ }\href@noop {} {\bibfield  {journal} {\bibinfo  {journal}
  {Physical Review}\ }\textbf {\bibinfo {volume} {62}},\ \bibinfo {pages} {438}
  (\bibinfo {year} {1942})}\BibitemShut {NoStop}%
\bibitem [{\citenamefont {MacDonald}(1964)}]{macdonald1964coulomb}%
  \BibitemOpen
  \bibfield  {author} {\bibinfo {author} {\bibfnamefont {N.}~\bibnamefont
  {MacDonald}},\ }\href@noop {} {\bibfield  {journal} {\bibinfo  {journal}
  {Physics Letters}\ }\textbf {\bibinfo {volume} {10}},\ \bibinfo {pages} {334}
  (\bibinfo {year} {1964})}\BibitemShut {NoStop}%
\bibitem [{\citenamefont {Nebel}\ and\ \citenamefont
  {Lin}(1967)}]{nebel1967second}%
  \BibitemOpen
  \bibfield  {author} {\bibinfo {author} {\bibfnamefont {H.}~\bibnamefont
  {Nebel}}\ and\ \bibinfo {author} {\bibfnamefont {D.~L.}\ \bibnamefont
  {Lin}},\ }\href@noop {} {\bibfield  {journal} {\bibinfo  {journal} {Physical
  Review}\ }\textbf {\bibinfo {volume} {156}},\ \bibinfo {pages} {1133}
  (\bibinfo {year} {1967})}\BibitemShut {NoStop}%
\bibitem [{\citenamefont {Douglas}\ and\ \citenamefont
  {MacDonald}(1967)}]{douglas1967coulomb}%
  \BibitemOpen
  \bibfield  {author} {\bibinfo {author} {\bibfnamefont {A.~C.}\ \bibnamefont
  {Douglas}}\ and\ \bibinfo {author} {\bibfnamefont {N.}~\bibnamefont
  {MacDonald}},\ }\href@noop {} {\bibfield  {journal} {\bibinfo  {journal}
  {Physics Letters B}\ }\textbf {\bibinfo {volume} {24}},\ \bibinfo {pages}
  {447} (\bibinfo {year} {1967})}\BibitemShut {NoStop}%
\bibitem [{\citenamefont {Danos}\ and\ \citenamefont
  {Greiner}(1964)}]{danos1964dynamic}%
  \BibitemOpen
  \bibfield  {author} {\bibinfo {author} {\bibfnamefont {M.}~\bibnamefont
  {Danos}}\ and\ \bibinfo {author} {\bibfnamefont {W.}~\bibnamefont
  {Greiner}},\ }\href@noop {} {\bibfield  {journal} {\bibinfo  {journal}
  {Physical Review}\ }\textbf {\bibinfo {volume} {134}},\ \bibinfo {pages}
  {B284} (\bibinfo {year} {1964})}\BibitemShut {NoStop}%
\bibitem [{\citenamefont {Weber}\ \emph {et~al.}(1966)\citenamefont {Weber},
  \citenamefont {Huber},\ and\ \citenamefont {Greiner}}]{weber1966theorie}%
  \BibitemOpen
  \bibfield  {author} {\bibinfo {author} {\bibfnamefont {H.~J.}\ \bibnamefont
  {Weber}}, \bibinfo {author} {\bibfnamefont {M.~G.}\ \bibnamefont {Huber}},\
  and\ \bibinfo {author} {\bibfnamefont {W.}~\bibnamefont {Greiner}},\
  }\href@noop {} {\bibfield  {journal} {\bibinfo  {journal} {Zeitschrift
  f{\"u}r Physik}\ }\textbf {\bibinfo {volume} {192}},\ \bibinfo {pages} {182}
  (\bibinfo {year} {1966})}\BibitemShut {NoStop}%
\bibitem [{\citenamefont {Huber}\ \emph {et~al.}(1967)\citenamefont {Huber},
  \citenamefont {Danos}, \citenamefont {Weber},\ and\ \citenamefont
  {Greiner}}]{huber1967collective}%
  \BibitemOpen
  \bibfield  {author} {\bibinfo {author} {\bibfnamefont {M.}~\bibnamefont
  {Huber}}, \bibinfo {author} {\bibfnamefont {M.}~\bibnamefont {Danos}},
  \bibinfo {author} {\bibfnamefont {H.}~\bibnamefont {Weber}},\ and\ \bibinfo
  {author} {\bibfnamefont {W.}~\bibnamefont {Greiner}},\ }\href@noop {}
  {\bibfield  {journal} {\bibinfo  {journal} {Physical Review}\ }\textbf
  {\bibinfo {volume} {155}},\ \bibinfo {pages} {1073} (\bibinfo {year}
  {1967})}\BibitemShut {NoStop}%
\bibitem [{\citenamefont {Levinger}(1957)}]{levinger1957migdal}%
  \BibitemOpen
  \bibfield  {author} {\bibinfo {author} {\bibfnamefont {J.}~\bibnamefont
  {Levinger}},\ }\href@noop {} {\bibfield  {journal} {\bibinfo  {journal}
  {Physical Review}\ }\textbf {\bibinfo {volume} {107}},\ \bibinfo {pages}
  {554} (\bibinfo {year} {1957})}\BibitemShut {NoStop}%
\bibitem [{\citenamefont {Orce}\ \emph {et~al.}(2012)\citenamefont {Orce},
  \citenamefont {Drake}, \citenamefont {Djongolov}, \citenamefont
  {Navr{\'a}til},\ and\ \citenamefont {{\it et al.}}}]{orce2012reorientation}%
  \BibitemOpen
  \bibfield  {author} {\bibinfo {author} {\bibfnamefont {J.~N.}\ \bibnamefont
  {Orce}}, \bibinfo {author} {\bibfnamefont {T.~E.}\ \bibnamefont {Drake}},
  \bibinfo {author} {\bibfnamefont {M.~K.}\ \bibnamefont {Djongolov}}, \bibinfo
  {author} {\bibfnamefont {P.}~\bibnamefont {Navr{\'a}til}},\ and\ \bibinfo
  {author} {\bibnamefont {{\it et al.}}},\ }\href@noop {} {\bibfield  {journal}
  {\bibinfo  {journal} {Physical Review C}\ }\textbf {\bibinfo {volume} {86}},\
  \bibinfo {pages} {041303} (\bibinfo {year} {2012})}\BibitemShut {NoStop}%
\bibitem [{\citenamefont {M{\"o}ller}\ \emph {et~al.}(1995)\citenamefont
  {M{\"o}ller}, \citenamefont {Nix} \emph {et~al.}}]{moller1995atomic}%
  \BibitemOpen
  \bibfield  {author} {\bibinfo {author} {\bibfnamefont {P.}~\bibnamefont
  {M{\"o}ller}}, \bibinfo {author} {\bibfnamefont {J.~R.}\ \bibnamefont {Nix}},
  \emph {et~al.},\ }\href@noop {} {\bibfield  {journal} {\bibinfo  {journal}
  {Atomic Data Nuclear Data Tables}\ }\textbf {\bibinfo {volume} {66}},\
  \bibinfo {pages} {131} (\bibinfo {year} {1995})}\BibitemShut {NoStop}%
\bibitem [{\citenamefont {Sonzogni}(2007)}]{sonzogni2007nndc}%
  \BibitemOpen
  \bibfield  {author} {\bibinfo {author} {\bibfnamefont {A.}~\bibnamefont
  {Sonzogni}},\ }in\ \href@noop {} {\emph {\bibinfo {booktitle} {International
  Conference on Nuclear Data for Science and Technology}}}\ (\bibinfo
  {organization} {EDP Sciences},\ \bibinfo {year} {2007})\ pp.\ \bibinfo
  {pages} {105--106}\BibitemShut {NoStop}%
\bibitem [{\citenamefont {H{\"a}usser}\ \emph {et~al.}(1972)\citenamefont
  {H{\"a}usser}, \citenamefont {McDonald}, \citenamefont {Alexander},
  \citenamefont {Ferguson},\ and\ \citenamefont {Warner}}]{hausser1972nuclear}%
  \BibitemOpen
  \bibfield  {author} {\bibinfo {author} {\bibfnamefont {O.}~\bibnamefont
  {H{\"a}usser}}, \bibinfo {author} {\bibfnamefont {A.~B.}\ \bibnamefont
  {McDonald}}, \bibinfo {author} {\bibfnamefont {T.~K.}\ \bibnamefont
  {Alexander}}, \bibinfo {author} {\bibfnamefont {A.~J.}\ \bibnamefont
  {Ferguson}},\ and\ \bibinfo {author} {\bibfnamefont {R.~E.}\ \bibnamefont
  {Warner}},\ }\href@noop {} {\bibfield  {journal} {\bibinfo  {journal}
  {Physics Letters B}\ }\textbf {\bibinfo {volume} {38}},\ \bibinfo {pages}
  {75} (\bibinfo {year} {1972})}\BibitemShut {NoStop}%
\bibitem [{\citenamefont {Disdier}\ \emph {et~al.}(1971)\citenamefont
  {Disdier}, \citenamefont {Ball}, \citenamefont {H{\"a}usser},\ and\
  \citenamefont {Warner}}]{disdier1971projectile}%
  \BibitemOpen
  \bibfield  {author} {\bibinfo {author} {\bibfnamefont {D.~L.}\ \bibnamefont
  {Disdier}}, \bibinfo {author} {\bibfnamefont {G.~C.}\ \bibnamefont {Ball}},
  \bibinfo {author} {\bibfnamefont {O.}~\bibnamefont {H{\"a}usser}},\ and\
  \bibinfo {author} {\bibfnamefont {R.~E.}\ \bibnamefont {Warner}},\
  }\href@noop {} {\bibfield  {journal} {\bibinfo  {journal} {Physical Review
  Letters}\ }\textbf {\bibinfo {volume} {27}},\ \bibinfo {pages} {1391}
  (\bibinfo {year} {1971})}\BibitemShut {NoStop}%
\bibitem [{\citenamefont {Vermeer}\ \emph {et~al.}(1982)\citenamefont
  {Vermeer}, \citenamefont {Zabel}, \citenamefont {Esat}, \citenamefont
  {Kuehner}, \citenamefont {Spear},\ and\ \citenamefont
  {Baxter}}]{vermeer1982giant}%
  \BibitemOpen
  \bibfield  {author} {\bibinfo {author} {\bibfnamefont {W.~J.}\ \bibnamefont
  {Vermeer}}, \bibinfo {author} {\bibfnamefont {T.~H.}\ \bibnamefont {Zabel}},
  \bibinfo {author} {\bibfnamefont {M.~T.}\ \bibnamefont {Esat}}, \bibinfo
  {author} {\bibfnamefont {J.~A.}\ \bibnamefont {Kuehner}}, \bibinfo {author}
  {\bibfnamefont {R.~H.}\ \bibnamefont {Spear}},\ and\ \bibinfo {author}
  {\bibfnamefont {A.~M.}\ \bibnamefont {Baxter}},\ }\href@noop {} {\bibfield
  {journal} {\bibinfo  {journal} {Australian Journal of Physics}\ }\textbf
  {\bibinfo {volume} {35}},\ \bibinfo {pages} {283} (\bibinfo {year}
  {1982})}\BibitemShut {NoStop}%
\bibitem [{\citenamefont {Vermeer}\ \emph {et~al.}(1983)\citenamefont
  {Vermeer}, \citenamefont {Esat}, \citenamefont {Kuehner}, \citenamefont
  {Spear}, \citenamefont {Baxter},\ and\ \citenamefont
  {Hinds}}]{vermeer1983electric}%
  \BibitemOpen
  \bibfield  {author} {\bibinfo {author} {\bibfnamefont {W.~J.}\ \bibnamefont
  {Vermeer}}, \bibinfo {author} {\bibfnamefont {M.~T.}\ \bibnamefont {Esat}},
  \bibinfo {author} {\bibfnamefont {J.~A.}\ \bibnamefont {Kuehner}}, \bibinfo
  {author} {\bibfnamefont {R.~H.}\ \bibnamefont {Spear}}, \bibinfo {author}
  {\bibfnamefont {A.~M.}\ \bibnamefont {Baxter}},\ and\ \bibinfo {author}
  {\bibfnamefont {S.}~\bibnamefont {Hinds}},\ }\href@noop {} {\bibfield
  {journal} {\bibinfo  {journal} {Physics Letters B}\ }\textbf {\bibinfo
  {volume} {122}},\ \bibinfo {pages} {23} (\bibinfo {year} {1983})}\BibitemShut
  {NoStop}%
\bibitem [{\citenamefont {Navr{\'a}til}(2007)}]{navratil2007local}%
  \BibitemOpen
  \bibfield  {author} {\bibinfo {author} {\bibfnamefont {P.}~\bibnamefont
  {Navr{\'a}til}},\ }\href@noop {} {\bibfield  {journal} {\bibinfo  {journal}
  {Few-Body Systems}\ }\textbf {\bibinfo {volume} {41}},\ \bibinfo {pages}
  {117} (\bibinfo {year} {2007})}\BibitemShut {NoStop}%
\bibitem [{\citenamefont {Roth}\ \emph {et~al.}(2014)\citenamefont {Roth},
  \citenamefont {Calci}, \citenamefont {Langhammer},\ and\ \citenamefont
  {Binder}}]{roth2014evolved}%
  \BibitemOpen
  \bibfield  {author} {\bibinfo {author} {\bibfnamefont {R.}~\bibnamefont
  {Roth}}, \bibinfo {author} {\bibfnamefont {A.}~\bibnamefont {Calci}},
  \bibinfo {author} {\bibfnamefont {J.}~\bibnamefont {Langhammer}},\ and\
  \bibinfo {author} {\bibfnamefont {S.}~\bibnamefont {Binder}},\ }\href@noop {}
  {\bibfield  {journal} {\bibinfo  {journal} {Physical Review C}\ }\textbf
  {\bibinfo {volume} {90}},\ \bibinfo {pages} {024325} (\bibinfo {year}
  {2014})}\BibitemShut {NoStop}%
\bibitem [{\citenamefont {Entem}\ and\ \citenamefont
  {Machleidt}(2003)}]{entem2003accurate}%
  \BibitemOpen
  \bibfield  {author} {\bibinfo {author} {\bibfnamefont {D.~R.}\ \bibnamefont
  {Entem}}\ and\ \bibinfo {author} {\bibfnamefont {R.}~\bibnamefont
  {Machleidt}},\ }\href@noop {} {\bibfield  {journal} {\bibinfo  {journal}
  {Physical Review C}\ }\textbf {\bibinfo {volume} {68}},\ \bibinfo {pages}
  {041001} (\bibinfo {year} {2003})}\BibitemShut {NoStop}%
\bibitem [{\citenamefont {Entem}\ \emph {et~al.}(2017)\citenamefont {Entem},
  \citenamefont {Machleidt},\ and\ \citenamefont {Nosyk}}]{entem2017high}%
  \BibitemOpen
  \bibfield  {author} {\bibinfo {author} {\bibfnamefont {D.~R.}\ \bibnamefont
  {Entem}}, \bibinfo {author} {\bibfnamefont {R.}~\bibnamefont {Machleidt}},\
  and\ \bibinfo {author} {\bibfnamefont {Y.}~\bibnamefont {Nosyk}},\
  }\href@noop {} {\bibfield  {journal} {\bibinfo  {journal} {Physical Review
  C}\ }\textbf {\bibinfo {volume} {96}},\ \bibinfo {pages} {024004} (\bibinfo
  {year} {2017})}\BibitemShut {NoStop}%
\bibitem [{\citenamefont {Entem}\ \emph {et~al.}(2015)\citenamefont {Entem},
  \citenamefont {Kaiser}, \citenamefont {Machleidt},\ and\ \citenamefont
  {Nosyk}}]{entem2015peripheral}%
  \BibitemOpen
  \bibfield  {author} {\bibinfo {author} {\bibfnamefont {D.~R.}\ \bibnamefont
  {Entem}}, \bibinfo {author} {\bibfnamefont {N.}~\bibnamefont {Kaiser}},
  \bibinfo {author} {\bibfnamefont {R.}~\bibnamefont {Machleidt}},\ and\
  \bibinfo {author} {\bibfnamefont {Y.}~\bibnamefont {Nosyk}},\ }\href@noop {}
  {\bibfield  {journal} {\bibinfo  {journal} {Physical Review C}\ }\textbf
  {\bibinfo {volume} {91}},\ \bibinfo {pages} {014002} (\bibinfo {year}
  {2015})}\BibitemShut {NoStop}%
\bibitem [{\citenamefont {Bogner}\ \emph {et~al.}(2007)\citenamefont {Bogner},
  \citenamefont {Furnstahl},\ and\ \citenamefont
  {Perry}}]{bogner2007similarity}%
  \BibitemOpen
  \bibfield  {author} {\bibinfo {author} {\bibfnamefont {S.~K.}\ \bibnamefont
  {Bogner}}, \bibinfo {author} {\bibfnamefont {R.~J.}\ \bibnamefont
  {Furnstahl}},\ and\ \bibinfo {author} {\bibfnamefont {R.~J.}\ \bibnamefont
  {Perry}},\ }\href@noop {} {\bibfield  {journal} {\bibinfo  {journal}
  {Physical Review C}\ }\textbf {\bibinfo {volume} {75}},\ \bibinfo {pages}
  {061001} (\bibinfo {year} {2007})}\BibitemShut {NoStop}%
\bibitem [{\citenamefont {Haydock}(1974)}]{haydock1974inverse}%
  \BibitemOpen
  \bibfield  {author} {\bibinfo {author} {\bibfnamefont {R.}~\bibnamefont
  {Haydock}},\ }\href@noop {} {\bibfield  {journal} {\bibinfo  {journal}
  {Journal of Physics A: Mathematical, Nuclear and General}\ }\textbf {\bibinfo
  {volume} {7}},\ \bibinfo {pages} {2120} (\bibinfo {year} {1974})}\BibitemShut
  {NoStop}%
\bibitem [{\citenamefont {Marchisio}\ \emph {et~al.}(2003)\citenamefont
  {Marchisio}, \citenamefont {Barnea}, \citenamefont {Leidemann},\ and\
  \citenamefont {Orlandini}}]{marchisio2003efficient}%
  \BibitemOpen
  \bibfield  {author} {\bibinfo {author} {\bibfnamefont {M.~A.}\ \bibnamefont
  {Marchisio}}, \bibinfo {author} {\bibfnamefont {N.}~\bibnamefont {Barnea}},
  \bibinfo {author} {\bibfnamefont {W.}~\bibnamefont {Leidemann}},\ and\
  \bibinfo {author} {\bibfnamefont {G.}~\bibnamefont {Orlandini}},\ }\href@noop
  {} {\bibfield  {journal} {\bibinfo  {journal} {Few-Body Systems}\ }\textbf
  {\bibinfo {volume} {33}},\ \bibinfo {pages} {259} (\bibinfo {year}
  {2003})}\BibitemShut {NoStop}%
\bibitem [{\citenamefont {Ngwetsheni}(2023)}]{cebotobe}%
  \BibitemOpen
  \bibfield  {author} {\bibinfo {author} {\bibfnamefont {C.~{\textit{et
  al.}}.}\ \bibnamefont {Ngwetsheni}},\ }\href@noop {} {\bibfield  {journal}
  {\bibinfo  {journal} {Physical Review}\ ,\ \bibinfo {pages} {in preparation}}
  (\bibinfo {year} {2023})}\BibitemShut {NoStop}%
\bibitem [{\citenamefont {Nathans}\ and\ \citenamefont
  {Halpern}(1953)}]{nathans1953excitation}%
  \BibitemOpen
  \bibfield  {author} {\bibinfo {author} {\bibfnamefont {R.}~\bibnamefont
  {Nathans}}\ and\ \bibinfo {author} {\bibfnamefont {J.}~\bibnamefont
  {Halpern}},\ }\href@noop {} {\bibfield  {journal} {\bibinfo  {journal}
  {Physical Review}\ }\textbf {\bibinfo {volume} {92}},\ \bibinfo {pages} {940}
  (\bibinfo {year} {1953})}\BibitemShut {NoStop}%
\bibitem [{\citenamefont {Fuller}(1985)}]{fuller1985photonuclear}%
  \BibitemOpen
  \bibfield  {author} {\bibinfo {author} {\bibfnamefont {E.~G.}\ \bibnamefont
  {Fuller}},\ }\href@noop {} {\bibfield  {journal} {\bibinfo  {journal}
  {Physics Reports}\ }\textbf {\bibinfo {volume} {127}},\ \bibinfo {pages}
  {185} (\bibinfo {year} {1985})}\BibitemShut {NoStop}%
\bibitem [{\citenamefont {Brown}\ \emph {et~al.}(1988)\citenamefont {Brown},
  \citenamefont {Etchegoyen}, \citenamefont {Rae},\ and\ \citenamefont
  {Godwin}}]{brown1988computer}%
  \BibitemOpen
  \bibfield  {author} {\bibinfo {author} {\bibfnamefont {B.~A.}\ \bibnamefont
  {Brown}}, \bibinfo {author} {\bibfnamefont {A.}~\bibnamefont {Etchegoyen}},
  \bibinfo {author} {\bibfnamefont {W.~D.~M.}\ \bibnamefont {Rae}},\ and\
  \bibinfo {author} {\bibfnamefont {N.~S.}\ \bibnamefont {Godwin}},\
  }\href@noop {} {\bibfield  {journal} {\bibinfo  {journal} {MSU-NSCL Report}\
  }\textbf {\bibinfo {volume} {524}} (\bibinfo {year} {1988})}\BibitemShut
  {NoStop}%
\bibitem [{\citenamefont {Warburton}\ and\ \citenamefont
  {Brown}(1992)}]{warburton1992effective}%
  \BibitemOpen
  \bibfield  {author} {\bibinfo {author} {\bibfnamefont {E.~K.}\ \bibnamefont
  {Warburton}}\ and\ \bibinfo {author} {\bibfnamefont {B.~A.}\ \bibnamefont
  {Brown}},\ }\href@noop {} {\bibfield  {journal} {\bibinfo  {journal}
  {Physical Review C}\ }\textbf {\bibinfo {volume} {46}},\ \bibinfo {pages}
  {923} (\bibinfo {year} {1992})}\BibitemShut {NoStop}%
\bibitem [{\citenamefont {Lubna}\ \emph {et~al.}(2019)\citenamefont {Lubna},
  \citenamefont {Kravvaris}, \citenamefont {Tabor}, \citenamefont {Tripathi},
  \citenamefont {Volya}, \citenamefont {Rubino}, \citenamefont {Allmond},
  \citenamefont {Abromeit}, \citenamefont {Baby},\ and\ \citenamefont
  {Hensley}}]{lubna2019structure}%
  \BibitemOpen
  \bibfield  {author} {\bibinfo {author} {\bibfnamefont {R.~S.}\ \bibnamefont
  {Lubna}}, \bibinfo {author} {\bibfnamefont {K.}~\bibnamefont {Kravvaris}},
  \bibinfo {author} {\bibfnamefont {S.~L.}\ \bibnamefont {Tabor}}, \bibinfo
  {author} {\bibfnamefont {V.}~\bibnamefont {Tripathi}}, \bibinfo {author}
  {\bibfnamefont {A.}~\bibnamefont {Volya}}, \bibinfo {author} {\bibfnamefont
  {E.}~\bibnamefont {Rubino}}, \bibinfo {author} {\bibfnamefont
  {J.}~\bibnamefont {Allmond}}, \bibinfo {author} {\bibfnamefont
  {B.}~\bibnamefont {Abromeit}}, \bibinfo {author} {\bibfnamefont
  {L.}~\bibnamefont {Baby}},\ and\ \bibinfo {author} {\bibfnamefont
  {T.}~\bibnamefont {Hensley}},\ }\href@noop {} {\bibfield  {journal} {\bibinfo
   {journal} {Physical Review C}\ }\textbf {\bibinfo {volume} {100}},\ \bibinfo
  {pages} {034308} (\bibinfo {year} {2019})}\BibitemShut {NoStop}%
\bibitem [{\citenamefont {Lubna}\ \emph {et~al.}(2020)\citenamefont {Lubna},
  \citenamefont {Kravvaris}, \citenamefont {Tabor}, \citenamefont {Tripathi},
  \citenamefont {Rubino},\ and\ \citenamefont {Volya}}]{lubna2020evolution}%
  \BibitemOpen
  \bibfield  {author} {\bibinfo {author} {\bibfnamefont {R.~S.}\ \bibnamefont
  {Lubna}}, \bibinfo {author} {\bibfnamefont {K.}~\bibnamefont {Kravvaris}},
  \bibinfo {author} {\bibfnamefont {S.~L.}\ \bibnamefont {Tabor}}, \bibinfo
  {author} {\bibfnamefont {V.}~\bibnamefont {Tripathi}}, \bibinfo {author}
  {\bibfnamefont {E.}~\bibnamefont {Rubino}},\ and\ \bibinfo {author}
  {\bibfnamefont {A.}~\bibnamefont {Volya}},\ }\href@noop {} {\bibfield
  {journal} {\bibinfo  {journal} {Physical Review Research}\ }\textbf {\bibinfo
  {volume} {2}},\ \bibinfo {pages} {043342} (\bibinfo {year}
  {2020})}\BibitemShut {NoStop}%
\bibitem [{\citenamefont {Brown}(2022)}]{brown2022nuclear}%
  \BibitemOpen
  \bibfield  {author} {\bibinfo {author} {\bibfnamefont {B.~A.}\ \bibnamefont
  {Brown}},\ }\href@noop {} {\bibfield  {journal} {\bibinfo  {journal}
  {Physics}\ }\textbf {\bibinfo {volume} {4}},\ \bibinfo {pages} {525}
  (\bibinfo {year} {2022})}\BibitemShut {NoStop}%
\bibitem [{\citenamefont {Bassel}\ \emph {et~al.}(1982)\citenamefont {Bassel},
  \citenamefont {Brown}, \citenamefont {Lindsay},\ and\ \citenamefont
  {Rowley}}]{bassel19820+}%
  \BibitemOpen
  \bibfield  {author} {\bibinfo {author} {\bibfnamefont {R.~H.}\ \bibnamefont
  {Bassel}}, \bibinfo {author} {\bibfnamefont {B.~A.}\ \bibnamefont {Brown}},
  \bibinfo {author} {\bibfnamefont {R.}~\bibnamefont {Lindsay}},\ and\ \bibinfo
  {author} {\bibfnamefont {N.}~\bibnamefont {Rowley}},\ }\href@noop {}
  {\bibfield  {journal} {\bibinfo  {journal} {Journal of Physics G: Nuclear
  Physics}\ }\textbf {\bibinfo {volume} {8}},\ \bibinfo {pages} {1215}
  (\bibinfo {year} {1982})}\BibitemShut {NoStop}%
\bibitem [{\citenamefont {Richter}\ \emph {et~al.}(2008)\citenamefont
  {Richter}, \citenamefont {Mkhize},\ and\ \citenamefont
  {Brown}}]{richter2008sd}%
  \BibitemOpen
  \bibfield  {author} {\bibinfo {author} {\bibfnamefont {W.}~\bibnamefont
  {Richter}}, \bibinfo {author} {\bibfnamefont {S.}~\bibnamefont {Mkhize}},\
  and\ \bibinfo {author} {\bibfnamefont {B.~A.}\ \bibnamefont {Brown}},\
  }\href@noop {} {\bibfield  {journal} {\bibinfo  {journal} {Physical Review
  C}\ }\textbf {\bibinfo {volume} {78}},\ \bibinfo {pages} {064302} (\bibinfo
  {year} {2008})}\BibitemShut {NoStop}%
\bibitem [{\citenamefont {Epelbaum}\ \emph {et~al.}(2012)\citenamefont
  {Epelbaum}, \citenamefont {Krebs}, \citenamefont {L{\"a}hde}, \citenamefont
  {Lee},\ and\ \citenamefont {Mei{\ss}ner}}]{epelbaum2012structure}%
  \BibitemOpen
  \bibfield  {author} {\bibinfo {author} {\bibfnamefont {E.}~\bibnamefont
  {Epelbaum}}, \bibinfo {author} {\bibfnamefont {H.}~\bibnamefont {Krebs}},
  \bibinfo {author} {\bibfnamefont {T.~A.}\ \bibnamefont {L{\"a}hde}}, \bibinfo
  {author} {\bibfnamefont {D.}~\bibnamefont {Lee}},\ and\ \bibinfo {author}
  {\bibfnamefont {U.-G.}\ \bibnamefont {Mei{\ss}ner}},\ }\href@noop {}
  {\bibfield  {journal} {\bibinfo  {journal} {Physical Review Letters}\
  }\textbf {\bibinfo {volume} {109}},\ \bibinfo {pages} {252501} (\bibinfo
  {year} {2012})}\BibitemShut {NoStop}%
\bibitem [{\citenamefont {Stroberg}\ \emph {et~al.}(2022)\citenamefont
  {Stroberg}, \citenamefont {Henderson}, \citenamefont {Hackman}, \citenamefont
  {Ruotsalainen}, \citenamefont {Hagen},\ and\ \citenamefont
  {Holt}}]{stroberg2022systematics}%
  \BibitemOpen
  \bibfield  {author} {\bibinfo {author} {\bibfnamefont {S.~R.}\ \bibnamefont
  {Stroberg}}, \bibinfo {author} {\bibfnamefont {J.}~\bibnamefont {Henderson}},
  \bibinfo {author} {\bibfnamefont {G.}~\bibnamefont {Hackman}}, \bibinfo
  {author} {\bibfnamefont {P.}~\bibnamefont {Ruotsalainen}}, \bibinfo {author}
  {\bibfnamefont {G.}~\bibnamefont {Hagen}},\ and\ \bibinfo {author}
  {\bibfnamefont {J.~D.}\ \bibnamefont {Holt}},\ }\href@noop {} {\bibfield
  {journal} {\bibinfo  {journal} {Physical Review C}\ }\textbf {\bibinfo
  {volume} {105}},\ \bibinfo {pages} {034333} (\bibinfo {year}
  {2022})}\BibitemShut {NoStop}%
\bibitem [{\citenamefont {Henderson}\ \emph {et~al.}(2022)\citenamefont
  {Henderson}, \citenamefont {Hackman}, \citenamefont {Ruotsalainen},
  \citenamefont {Holt}, \citenamefont {Stroberg}, \citenamefont {Andreoiu},
  \citenamefont {Ball}, \citenamefont {Bernier}, \citenamefont {Bowry},
  \citenamefont {Caballero-Folch} \emph {et~al.}}]{henderson2022coulomb}%
  \BibitemOpen
  \bibfield  {author} {\bibinfo {author} {\bibfnamefont {J.}~\bibnamefont
  {Henderson}}, \bibinfo {author} {\bibfnamefont {G.}~\bibnamefont {Hackman}},
  \bibinfo {author} {\bibfnamefont {P.}~\bibnamefont {Ruotsalainen}}, \bibinfo
  {author} {\bibfnamefont {J.~D.}\ \bibnamefont {Holt}}, \bibinfo {author}
  {\bibfnamefont {S.~R.}\ \bibnamefont {Stroberg}}, \bibinfo {author}
  {\bibfnamefont {C.}~\bibnamefont {Andreoiu}}, \bibinfo {author}
  {\bibfnamefont {G.~C.}\ \bibnamefont {Ball}}, \bibinfo {author}
  {\bibfnamefont {N.}~\bibnamefont {Bernier}}, \bibinfo {author} {\bibfnamefont
  {M.}~\bibnamefont {Bowry}}, \bibinfo {author} {\bibfnamefont
  {R.}~\bibnamefont {Caballero-Folch}}, \emph {et~al.},\ }\href@noop {}
  {\bibfield  {journal} {\bibinfo  {journal} {Physical Review C}\ }\textbf
  {\bibinfo {volume} {105}},\ \bibinfo {pages} {034332} (\bibinfo {year}
  {2022})}\BibitemShut {NoStop}%
\bibitem [{\citenamefont {Sarma}\ and\ \citenamefont
  {Srivastava}(2023)}]{sarma2023ab}%
  \BibitemOpen
  \bibfield  {author} {\bibinfo {author} {\bibfnamefont {C.}~\bibnamefont
  {Sarma}}\ and\ \bibinfo {author} {\bibfnamefont {P.~C.}\ \bibnamefont
  {Srivastava}},\ }\href@noop {} {\bibfield  {journal} {\bibinfo  {journal}
  {Journal of Physics G: Nuclear and Particle Physics}\ }\textbf {\bibinfo
  {volume} {50}},\ \bibinfo {pages} {045105} (\bibinfo {year}
  {2023})}\BibitemShut {NoStop}%
\bibitem [{\citenamefont {Pritychenko}\ \emph {et~al.}(2016)\citenamefont
  {Pritychenko}, \citenamefont {Birch}, \citenamefont {Singh},\ and\
  \citenamefont {Horoi}}]{pritychenko2016tables}%
  \BibitemOpen
  \bibfield  {author} {\bibinfo {author} {\bibfnamefont {B.}~\bibnamefont
  {Pritychenko}}, \bibinfo {author} {\bibfnamefont {M.}~\bibnamefont {Birch}},
  \bibinfo {author} {\bibfnamefont {B.}~\bibnamefont {Singh}},\ and\ \bibinfo
  {author} {\bibfnamefont {M.}~\bibnamefont {Horoi}},\ }\href@noop {}
  {\bibfield  {journal} {\bibinfo  {journal} {Atomic Data and Nuclear Data
  Tables}\ }\textbf {\bibinfo {volume} {107}},\ \bibinfo {pages} {1} (\bibinfo
  {year} {2016})}\BibitemShut {NoStop}%
\bibitem [{\citenamefont {Brown}(2017)}]{brown2017oxygen}%
  \BibitemOpen
  \bibfield  {author} {\bibinfo {author} {\bibfnamefont {B.~A.}\ \bibnamefont
  {Brown}},\ }\href@noop {} {\bibfield  {journal} {\bibinfo  {journal}
  {International Journal of Modern Physics E}\ }\textbf {\bibinfo {volume}
  {26}},\ \bibinfo {pages} {1740003} (\bibinfo {year} {2017})}\BibitemShut
  {NoStop}%
\bibitem [{\citenamefont {Lawson}\ \emph {et~al.}(1976)\citenamefont {Lawson},
  \citenamefont {Serduke},\ and\ \citenamefont
  {Fortune}}]{lawson1976structure}%
  \BibitemOpen
  \bibfield  {author} {\bibinfo {author} {\bibfnamefont {R.~L.}\ \bibnamefont
  {Lawson}}, \bibinfo {author} {\bibfnamefont {F.~J.~D.}\ \bibnamefont
  {Serduke}},\ and\ \bibinfo {author} {\bibfnamefont {H.~T.}\ \bibnamefont
  {Fortune}},\ }\href@noop {} {\bibfield  {journal} {\bibinfo  {journal}
  {Physical Review C}\ }\textbf {\bibinfo {volume} {14}},\ \bibinfo {pages}
  {1245} (\bibinfo {year} {1976})}\BibitemShut {NoStop}%
\bibitem [{\citenamefont {Ellis}\ and\ \citenamefont
  {Engeland}(1970)}]{ellis1970weak}%
  \BibitemOpen
  \bibfield  {author} {\bibinfo {author} {\bibfnamefont {P.~J.}\ \bibnamefont
  {Ellis}}\ and\ \bibinfo {author} {\bibfnamefont {T.}~\bibnamefont
  {Engeland}},\ }\href@noop {} {\bibfield  {journal} {\bibinfo  {journal}
  {Nuclear Physics A}\ }\textbf {\bibinfo {volume} {144}},\ \bibinfo {pages}
  {161} (\bibinfo {year} {1970})}\BibitemShut {NoStop}%
\bibitem [{\citenamefont {Engeland}\ and\ \citenamefont
  {Ellis}(1972)}]{engeland1972weak}%
  \BibitemOpen
  \bibfield  {author} {\bibinfo {author} {\bibfnamefont {T.}~\bibnamefont
  {Engeland}}\ and\ \bibinfo {author} {\bibfnamefont {P.~J.}\ \bibnamefont
  {Ellis}},\ }\href@noop {} {\bibfield  {journal} {\bibinfo  {journal} {Nuclear
  Physics A}\ }\textbf {\bibinfo {volume} {181}},\ \bibinfo {pages} {368}
  (\bibinfo {year} {1972})}\BibitemShut {NoStop}%
\bibitem [{\citenamefont {Cline}\ \emph {et~al.}(2012)\citenamefont {Cline},
  \citenamefont {Czosnyka}, \citenamefont {Hayes}, \citenamefont
  {Napiorkowski}, \citenamefont {Warr},\ and\ \citenamefont
  {Wu}}]{cline2012gosia}%
  \BibitemOpen
  \bibfield  {author} {\bibinfo {author} {\bibfnamefont {D.}~\bibnamefont
  {Cline}}, \bibinfo {author} {\bibfnamefont {T.}~\bibnamefont {Czosnyka}},
  \bibinfo {author} {\bibfnamefont {A.~B.}\ \bibnamefont {Hayes}}, \bibinfo
  {author} {\bibfnamefont {P.}~\bibnamefont {Napiorkowski}}, \bibinfo {author}
  {\bibfnamefont {N.}~\bibnamefont {Warr}},\ and\ \bibinfo {author}
  {\bibfnamefont {C.~Y.}\ \bibnamefont {Wu}},\ }\href@noop {} {\bibfield
  {journal} {\bibinfo  {journal} {Gosia Steering Committee}\ }\textbf {\bibinfo
  {volume} {18}},\ \bibinfo {pages} {19} (\bibinfo {year} {2012})}\BibitemShut
  {NoStop}%
\bibitem [{\citenamefont {Spear}(1981)}]{spear1981static}%
  \BibitemOpen
  \bibfield  {author} {\bibinfo {author} {\bibfnamefont {R.~H.}\ \bibnamefont
  {Spear}},\ }\href@noop {} {\bibfield  {journal} {\bibinfo  {journal} {Physics
  Reports}\ }\textbf {\bibinfo {volume} {73}},\ \bibinfo {pages} {369}
  (\bibinfo {year} {1981})}\BibitemShut {NoStop}%
\bibitem [{\citenamefont {Barker}(1984)}]{barker1984decay}%
  \BibitemOpen
  \bibfield  {author} {\bibinfo {author} {\bibfnamefont {F.~C.}\ \bibnamefont
  {Barker}},\ }\href@noop {} {\bibfield  {journal} {\bibinfo  {journal}
  {Australian Journal of Physics}\ }\textbf {\bibinfo {volume} {37}},\ \bibinfo
  {pages} {267} (\bibinfo {year} {1984})}\BibitemShut {NoStop}%
\bibitem [{\citenamefont {Nakayama}\ \emph {et~al.}(2001)\citenamefont
  {Nakayama}, \citenamefont {Yamagata}, \citenamefont {Akimune}, \citenamefont
  {Daito}, \citenamefont {Fujimura}, \citenamefont {Fujita}, \citenamefont
  {Fujiwara}, \citenamefont {Fushimi}, \citenamefont {Greenfield},
  \citenamefont {Kohri} \emph {et~al.}}]{nakayama2001dipole}%
  \BibitemOpen
  \bibfield  {author} {\bibinfo {author} {\bibfnamefont {S.}~\bibnamefont
  {Nakayama}}, \bibinfo {author} {\bibfnamefont {T.}~\bibnamefont {Yamagata}},
  \bibinfo {author} {\bibfnamefont {H.}~\bibnamefont {Akimune}}, \bibinfo
  {author} {\bibfnamefont {I.}~\bibnamefont {Daito}}, \bibinfo {author}
  {\bibfnamefont {H.}~\bibnamefont {Fujimura}}, \bibinfo {author}
  {\bibfnamefont {Y.}~\bibnamefont {Fujita}}, \bibinfo {author} {\bibfnamefont
  {M.}~\bibnamefont {Fujiwara}}, \bibinfo {author} {\bibfnamefont
  {K.}~\bibnamefont {Fushimi}}, \bibinfo {author} {\bibfnamefont {M.~B.}\
  \bibnamefont {Greenfield}}, \bibinfo {author} {\bibfnamefont
  {H.}~\bibnamefont {Kohri}}, \emph {et~al.},\ }\href@noop {} {\bibfield
  {journal} {\bibinfo  {journal} {Physical Review Letters}\ }\textbf {\bibinfo
  {volume} {87}},\ \bibinfo {pages} {122502} (\bibinfo {year}
  {2001})}\BibitemShut {NoStop}%
\bibitem [{\citenamefont {Burda}\ \emph {et~al.}(2010)\citenamefont {Burda},
  \citenamefont {von Neumann-Cosel}, \citenamefont {Richter}, \citenamefont
  {Forss{\'e}n},\ and\ \citenamefont {Brown}}]{burda2010resonance}%
  \BibitemOpen
  \bibfield  {author} {\bibinfo {author} {\bibfnamefont {O.}~\bibnamefont
  {Burda}}, \bibinfo {author} {\bibfnamefont {P.}~\bibnamefont {von
  Neumann-Cosel}}, \bibinfo {author} {\bibfnamefont {A.}~\bibnamefont
  {Richter}}, \bibinfo {author} {\bibfnamefont {C.}~\bibnamefont
  {Forss{\'e}n}},\ and\ \bibinfo {author} {\bibfnamefont {B.~A.}\ \bibnamefont
  {Brown}},\ }\href@noop {} {\bibfield  {journal} {\bibinfo  {journal}
  {Physical Review C}\ }\textbf {\bibinfo {volume} {82}},\ \bibinfo {pages}
  {015808} (\bibinfo {year} {2010})}\BibitemShut {NoStop}%
\bibitem [{\citenamefont {Eramzhyan}\ \emph {et~al.}(1986)\citenamefont
  {Eramzhyan}, \citenamefont {Ishkhanov}, \citenamefont {Kapitonov},\ and\
  \citenamefont {Neudatchin}}]{eramzhyan1986giant}%
  \BibitemOpen
  \bibfield  {author} {\bibinfo {author} {\bibfnamefont {R.~A.}\ \bibnamefont
  {Eramzhyan}}, \bibinfo {author} {\bibfnamefont {B.~S.}\ \bibnamefont
  {Ishkhanov}}, \bibinfo {author} {\bibfnamefont {I.~M.}\ \bibnamefont
  {Kapitonov}},\ and\ \bibinfo {author} {\bibfnamefont {V.~G.}\ \bibnamefont
  {Neudatchin}},\ }\href@noop {} {\bibfield  {journal} {\bibinfo  {journal}
  {Physics Reports}\ }\textbf {\bibinfo {volume} {136}},\ \bibinfo {pages}
  {229} (\bibinfo {year} {1986})}\BibitemShut {NoStop}%
\bibitem [{\citenamefont {Neudatchin}\ \emph {et~al.}(1979)\citenamefont
  {Neudatchin}, \citenamefont {Smirnov},\ and\ \citenamefont
  {Golovanova}}]{neudatchin1979clustering}%
  \BibitemOpen
  \bibfield  {author} {\bibinfo {author} {\bibfnamefont {V.~G.}\ \bibnamefont
  {Neudatchin}}, \bibinfo {author} {\bibfnamefont {Y.~F.}\ \bibnamefont
  {Smirnov}},\ and\ \bibinfo {author} {\bibfnamefont {N.~F.}\ \bibnamefont
  {Golovanova}},\ }\href@noop {} {\bibfield  {journal} {\bibinfo  {journal}
  {Adv. Nucl. Phys.;(United States)}\ }\textbf {\bibinfo {volume} {11}}
  (\bibinfo {year} {1979})}\BibitemShut {NoStop}%
\bibitem [{\citenamefont {He}\ \emph {et~al.}(2014)\citenamefont {He},
  \citenamefont {Ma}, \citenamefont {Cao}, \citenamefont {Cai}, \citenamefont
  {Zhang} \emph {et~al.}}]{he2014giant}%
  \BibitemOpen
  \bibfield  {author} {\bibinfo {author} {\bibfnamefont {W.~B.}\ \bibnamefont
  {He}}, \bibinfo {author} {\bibfnamefont {Y.~G.}\ \bibnamefont {Ma}}, \bibinfo
  {author} {\bibfnamefont {X.~G.}\ \bibnamefont {Cao}}, \bibinfo {author}
  {\bibfnamefont {X.~Z.}\ \bibnamefont {Cai}}, \bibinfo {author} {\bibfnamefont
  {G.~Q.}\ \bibnamefont {Zhang}}, \emph {et~al.},\ }\href@noop {} {\bibfield
  {journal} {\bibinfo  {journal} {Physical Review Letters}\ }\textbf {\bibinfo
  {volume} {113}},\ \bibinfo {pages} {032506} (\bibinfo {year}
  {2014})}\BibitemShut {NoStop}%
\bibitem [{\citenamefont {Smilansky}\ \emph {et~al.}(1972)\citenamefont
  {Smilansky}, \citenamefont {Povh},\ and\ \citenamefont
  {Traxel}}]{smilansky1972role}%
  \BibitemOpen
  \bibfield  {author} {\bibinfo {author} {\bibfnamefont {U.}~\bibnamefont
  {Smilansky}}, \bibinfo {author} {\bibfnamefont {B.}~\bibnamefont {Povh}},\
  and\ \bibinfo {author} {\bibfnamefont {K.}~\bibnamefont {Traxel}},\
  }\href@noop {} {\bibfield  {journal} {\bibinfo  {journal} {Physics Letters
  B}\ }\textbf {\bibinfo {volume} {38}},\ \bibinfo {pages} {293} (\bibinfo
  {year} {1972})}\BibitemShut {NoStop}%
\bibitem [{\citenamefont {Weller}\ \emph {et~al.}(1985)\citenamefont {Weller},
  \citenamefont {Egelhof}, \citenamefont {{\v{C}}aplar}, \citenamefont
  {Karban}, \citenamefont {Kr{\"a}mer}, \citenamefont {M{\"o}bius},
  \citenamefont {Moroz}, \citenamefont {Rusek}, \citenamefont {Steffens},
  \citenamefont {Tungate} \emph {et~al.}}]{weller1985electromagnetic}%
  \BibitemOpen
  \bibfield  {author} {\bibinfo {author} {\bibfnamefont {A.}~\bibnamefont
  {Weller}}, \bibinfo {author} {\bibfnamefont {P.}~\bibnamefont {Egelhof}},
  \bibinfo {author} {\bibfnamefont {R.}~\bibnamefont {{\v{C}}aplar}}, \bibinfo
  {author} {\bibfnamefont {O.}~\bibnamefont {Karban}}, \bibinfo {author}
  {\bibfnamefont {D.}~\bibnamefont {Kr{\"a}mer}}, \bibinfo {author}
  {\bibfnamefont {K.-H.}\ \bibnamefont {M{\"o}bius}}, \bibinfo {author}
  {\bibfnamefont {Z.}~\bibnamefont {Moroz}}, \bibinfo {author} {\bibfnamefont
  {K.}~\bibnamefont {Rusek}}, \bibinfo {author} {\bibfnamefont
  {E.}~\bibnamefont {Steffens}}, \bibinfo {author} {\bibfnamefont
  {G.}~\bibnamefont {Tungate}}, \emph {et~al.},\ }\href@noop {} {\bibfield
  {journal} {\bibinfo  {journal} {Physical Review Letters}\ }\textbf {\bibinfo
  {volume} {55}},\ \bibinfo {pages} {480} (\bibinfo {year} {1985})}\BibitemShut
  {NoStop}%
\bibitem [{\citenamefont {Ball}\ \emph {et~al.}(1982)\citenamefont {Ball},
  \citenamefont {Alexander}, \citenamefont {Davies}, \citenamefont {Forster},\
  and\ \citenamefont {Mitchell}}]{ball1982dbla}%
  \BibitemOpen
  \bibfield  {author} {\bibinfo {author} {\bibfnamefont {G.~C.}\ \bibnamefont
  {Ball}}, \bibinfo {author} {\bibfnamefont {T.~K.}\ \bibnamefont {Alexander}},
  \bibinfo {author} {\bibfnamefont {W.~G.}\ \bibnamefont {Davies}}, \bibinfo
  {author} {\bibfnamefont {J.~S.}\ \bibnamefont {Forster}},\ and\ \bibinfo
  {author} {\bibfnamefont {I.~V.}\ \bibnamefont {Mitchell}},\ }\href@noop {}
  {\bibfield  {journal} {\bibinfo  {journal} {Nuclear Physics A}\ }\textbf
  {\bibinfo {volume} {377}},\ \bibinfo {pages} {268} (\bibinfo {year}
  {1982})}\BibitemShut {NoStop}%
\bibitem [{\citenamefont {Raman}\ \emph {et~al.}(1987)\citenamefont {Raman},
  \citenamefont {Malarkey}, \citenamefont {Milner}, \citenamefont {Nestor~Jr},\
  and\ \citenamefont {Stelson}}]{raman1987transition}%
  \BibitemOpen
  \bibfield  {author} {\bibinfo {author} {\bibfnamefont {S.}~\bibnamefont
  {Raman}}, \bibinfo {author} {\bibfnamefont {C.~H.}\ \bibnamefont {Malarkey}},
  \bibinfo {author} {\bibfnamefont {W.}~\bibnamefont {Milner}}, \bibinfo
  {author} {\bibfnamefont {C.}~\bibnamefont {Nestor~Jr}},\ and\ \bibinfo
  {author} {\bibfnamefont {P.}~\bibnamefont {Stelson}},\ }\href@noop {}
  {\bibfield  {journal} {\bibinfo  {journal} {Atomic Data and Nuclear data
  tables}\ }\textbf {\bibinfo {volume} {36}},\ \bibinfo {pages} {1} (\bibinfo
  {year} {1987})}\BibitemShut {NoStop}%
\bibitem [{\citenamefont {Orce}\ \emph {et~al.}(2021)\citenamefont {Orce},
  \citenamefont {Martin-Montes}, \citenamefont {Abrahams}, \citenamefont
  {Ngwetsheni}, \citenamefont {Brown}, \citenamefont {Kumar-Raju},
  \citenamefont {Mehl}, \citenamefont {Mokgolobotho}, \citenamefont {Akakpo},
  \citenamefont {Mavela} \emph {et~al.}}]{orce2021reorientation}%
  \BibitemOpen
  \bibfield  {author} {\bibinfo {author} {\bibfnamefont {J.~N.}\ \bibnamefont
  {Orce}}, \bibinfo {author} {\bibfnamefont {E.~J.}\ \bibnamefont
  {Martin-Montes}}, \bibinfo {author} {\bibfnamefont {K.~J.}\ \bibnamefont
  {Abrahams}}, \bibinfo {author} {\bibfnamefont {C.}~\bibnamefont
  {Ngwetsheni}}, \bibinfo {author} {\bibfnamefont {B.~A.}\ \bibnamefont
  {Brown}}, \bibinfo {author} {\bibfnamefont {M.}~\bibnamefont {Kumar-Raju}},
  \bibinfo {author} {\bibfnamefont {C.~V.}\ \bibnamefont {Mehl}}, \bibinfo
  {author} {\bibfnamefont {M.~J.}\ \bibnamefont {Mokgolobotho}}, \bibinfo
  {author} {\bibfnamefont {E.~H.}\ \bibnamefont {Akakpo}}, \bibinfo {author}
  {\bibfnamefont {D.~L.}\ \bibnamefont {Mavela}}, \emph {et~al.},\ }\href@noop
  {} {\bibfield  {journal} {\bibinfo  {journal} {Physical Review C}\ }\textbf
  {\bibinfo {volume} {104}},\ \bibinfo {pages} {L061305} (\bibinfo {year}
  {2021})}\BibitemShut {NoStop}%
\end{thebibliography}%

\end{document}